\documentclass[fleqn,usenatbib]{mnras}

\usepackage{mathptmx}

\usepackage[T1]{fontenc}

\DeclareRobustCommand{\VAN}[3]{#2}
\let\VANthebibliography\thebibliography
\def\thebibliography{\DeclareRobustCommand{\VAN}[3]{##3}\VANthebibliography}

\usepackage{graphicx}	
\usepackage{amsmath}	
\usepackage{amssymb}	

\newcommand{\fpl}{\bar{n}_\mathrm{p}}

\newcommand{\Ntotpl}{N_\mathrm{p}}
\newcommand{\Neffn}{K_\mathrm{n}}
\newcommand{\NeffF}{K_\mathrm{F}}

\newcommand{\NtotplHost}{N_\mathrm{h}}
\newcommand{\FplH}{F_\mathrm{h}}

\title[Exoplanets in the Galactic context]{Exoplanets in the Galactic context: Planet occurrence rates in the thin disk, thick disk and stellar halo of \textit{Kepler} stars}

\author[D. Bashi \& S. Zucker]{
Dolev Bashi,$^{1}$\thanks{E-mail: dolevbashi@gmail.com}
Shay Zucker$^{1}$
\\
$^{1}$Porter School of the Environment and Earth Sciences, Raymond and Beverly Sackler Faculty of Exact Sciences, Tel Aviv University,Tel Aviv, 6997801, Israel\\
}

\date{Accepted 2021 December 6. Received 2021 November 7; in original form 2021 July 31}

\pubyear{2015}

\begin{document}
\label{firstpage}
\pagerange{\pageref{firstpage}--\pageref{lastpage}}
\maketitle

\begin{abstract}
In order to gain a better understanding of planet formation and evolution, it is important to examine the statistics of exoplanets in the Galactic context. By combining information on stellar elemental abundances and kinematics, we constructed separate samples of \textit{Kepler} stars according to their affiliation to the Galactic components of thin disk, thick disk and stellar halo. Using a Bayesian analysis with conjugate priors, we then investigated how planet occurrence rates differ in different regions of planet properties. We find that young, slow and metal-rich stars, associated mainly with the thin disk, host on average more planets (especially close-in super Earths) compared to the old, fast and metal-poor thick disk stars. We further assess the dependence between stellar properties such as spectral type and metallicity, and planet occurrence rates. The trends we find agree with those found by other authors as well. We argue that in the Galactic context, these are probably not the main properties that affect planet occurrence rates, but rather the dynamical history of stars, and especially stellar age and kinematics, impact the current distribution of planets in the Galaxy. 
\end{abstract}




\section{Introduction}
Studying the dependence between the properties of exoplanets and the properties of their host stars may contribute to a better understanding of the mechanisms underlying planet formation and evolution. The discoveries made by the \textit{Kepler} mission \citep{Boruckietal10} of thousands of exoplanets in multi-planet systems, allowed a more detailed characterisation of the planet population within a range of stellar types and properties. Early works by \cite{Howardetal12} and later by \cite{Muldersetal15} have already suggested that the occurrence rates of close-in super Earths and sub-Neptunes 
increase towards later spectral type at all separations. In more recent works, both \cite{Heetal20} and \cite{Yangetal20} have revisited the estimates of the fraction of planet host systems as a function of stellar mass and effective temperature. Both studies suggested that the fraction of stars with planets changes from $\sim 0.3 $ (in F-type stars) to $\sim 0.8 $ (in K-type stars) as a function of spectral type, while planet multiplicity decreases with spectral type (i.e. increasing stellar temperature and mass). Similarly, works focusing on stellar metallicity such as those of \cite{Buchhaveetal12} and \cite{Petiguraetal18}, have argued that metal-rich stars may host a greater diversity of planetary systems.

In the Galactic context, it is currently not fully understood how the structure and evolution of our Galaxy characterised by stellar birthplace, chemical compositions and dynamics may affect the possibility to form and maintain planets. In the solar neighbourhood, it is common to group stars into three main Galactic populations that were formerly labeled as: the thin and thick disks, and the stellar halo \citep{GilmoreReid83}. 
Over the past few decades further analysis of these populations based on spectroscopic surveys demonstrated that these populations of stars differ in kinematics (i.e. the positions and motions of the stars in the Galaxy), chemical composition and stellar age. As compared to thin disk stars, thick disk stars are consistently older, kinematically hotter, and more iron-poor and $\alpha$-enhanced \citep[e.g.][]{GilmoreReid83,Fuhrmann98, Bensbyetal03,Reddyetal06,Adibekyanetal12,Bensbyetal14,BashiZucker19,Bashietal20}. Furthermore, halo stars are believed to be the most ancient and metal-poor stars in our galaxy. The formation of the stellar halo is still not fully understood. The two main models proposed to explain the high-velocity dispersion of these stars are of 'accretion' and 'in-situ' scenarios \citep{NissenSchuster10}. The first pertains to stars that have been accreted through the mergers
of dwarf spheroidal-like galaxies such as "Gaia Sausage" \citep{Belokurovetal18} and "Gaia Enceladus" \citep{Helmietal18}, while the second suggests formation inside the galaxy and "kicking-out" from the disk \citep[][and references therein]{Conroyetal19}. Using elemental abundances, kinematics and age estimates, it is possible to distinguish in the solar neighbourhood between these populations. 

Attempts to put our knowledge about exoplanets in the Galactic context are still sporadic, as the vast majority of currently known planet-host stars are affiliated mainly with the thin disk, with only few thick-disk and halo planet-host stars. This should not be of any surprise as most stars in the solar neighbourhood belong to the thin disk. \cite{Ecuvillonetal07} were among the first to explore the kinematics of stars with and without planets. However, since their survey focused on iron-rich stars, the sample of thick-disk or halo stars was too small to draw any decisive conclusions. Later works of \cite{Haywood09} and \cite{Adibekyanetal12} explored the $\alpha$ enhancement of metal-poor planet-hosting stars and have shown that at low metallicities, planet-hosting stars tend to have high $\alpha$ content.
However, these studies did not take into account observational biases in their estimates of planet occurrence rates.

In a more recent work \citep{BashiZucker19}, we found that close-in super Earths and sub-Neptunes ($1-4R_\oplus$ , $P < 100$ days) orbiting stars with lower iron abundance ([Fe/H]$\sim-0.5$) and higher total speed tend to have a higher planet occurrence rate. On the other hand, we also noted that with fast stars having higher or much lower iron contents, planet occurrence rate decreases. Similar suggestive trends have also been demonstrated by \cite{Daietal21} who after attempting to correct for the influence of stellar properties on planet occurrence rate, they found that fast stars, based on stellar tangential velocity in the Galaxy, have a lower occurrence rate of close-in super Earths and sub-Neptunes as compared to slow stars. In a later work based on HARPS data \citep{Bashietal20}, we have found that the occurrence rates of low mass planets ($M_\mathrm{p}=1-20\,M_\oplus$, $P < 100$ days) in samples of thin and thick disk stars were comparable. To distinguish between groups, we used elemental abundance analysis.

In the current study, we aim to move one step forward in this direction, and examine planet occurrence rates in a well-defined sample of \textit{Kepler} FGK dwarfs, in their Galactic context. In the literature, there is currently no exact method to affiliate stars of the solar neighborhood to either the thin diske, thick disk or the stellar halo. Often, chemical composition criteria are used, but classifications based on kinematics, stellar age, spatial position or a combination thereof can also be found \citep{Bensbyetal03,Adibekyanetal12,Bashietal20}. It is likely that identifying a single population purely on the basis of a certain specific criterion would result in its contamination with stars of other Galactic components \citep{Bensbyetal14}. Therefore, in this work, we will define our sample by combining information on both kinematics and element abundance to produce carefully vetted samples of stars in the thin disk, thick disk and stellar halo. Our purpose is to check quantitatively whether there is a difference in the population of planets orbiting stars of different Galactic components and to further investigate the impact of the individual stellar properties (e.g. size, temperature and iron content) as compared to their Galactic context properties (kinematics, age and $\alpha$-enhancement) in determining planet occurrence rates.

The following section describes the way we compiled our samples of stars and planets. Section~\ref{Occurrence} presents the method we used to calculate planet occurrence rates. We present our results in Section~\ref{Results} and discuss our findings in Section~\ref{Discussion}.

\section{The sample}
\label{Thesample}
\subsection{Overview}

In order to construct samples of stars and planet-host stars separated into Galactic populations, we searched for \textit{Kepler} stars with reported information on both kinematics and elemental abundance. As we have discussed elsewhere \citep[][and references therein]{BashiZucker19}, stellar kinematics alone is not the ideal diagnostic to differentiate among the Galactic populations. However, it can be used in combination with information on element abundances ($\mathrm{[Fe/H]}$ and $\mathrm{[\alpha/Fe]}$).

The largest sample that can be compiled based on the available information about kinematics and elemental abundance can be compiled by first combining the five-parameter astrometric solution of \textit{Gaia} EDR3 \citep{GaiaCol21} with the radial velocities of the Large Sky Area Multi-Object Fibre Spectroscopic Telescope (LAMOST) DR5 in order to characterize the stellar kinematics. Thus one can obtain full space motions for a large sample of \textit{Kepler} stars. 

Next, we need to address the issue of $\mathrm{[Fe/H]}$ and $\mathrm{[\alpha/Fe]}$, which are usually used to distinguish between thin and thick disk stars \citep{Adibekyanetal12, Bensbyetal14, Bashietal20}. There have been some sporadic attempts in the past to estimate chemical abundances for a fraction of the \textit{Kepler} stars, most of them focused on confirmed planet-host stars and \textit{Kepler} objects of interests (KOIs). Those attempts included the Apache Point Observatory Galactic Evolution Experiment
\citep[APOGEE;][]{Majewskietal17} with its high resolution ($R \sim 22500$) spectrograph, and the California Kepler Survey \citep[CKS;][]{Petiguraetal17} using HIRES ($R \sim 85000$). However, the overall number of \textit{Kepler} stars that have chemical abundance estimates from high-precision spectroscopy is relatively small and biased towards planet-host stars, and therefore not suitable for planet occurrence-rate studies. 

We therefore decided to use for that purpose the catalogue of \cite{Xiangetal19}, who used a data-driven model ('DD-Payne') trained with data from the Galactic Archaeology with HERMES (GALAH) DR2 \citep{Buderetal18} and APOGEE DR14, and provided abundance estimates of $16$ elements for more than six million LAMOST DR5 stars. They demonstrated that it was possible to obtain precise elemental abundances from low-resolution spectra.

\subsection{Stellar sample}

We began by cross matching the LAMOST DR5 DD-Payne sample \citep{Xiangetal19} with the \textit{Kepler} input catalog (KIC) \citep{Mathuretal17} and \textit{Gaia} EDR3 \citep{GaiaCol21}. The cross-match was based on angular distance smaller than $1\,\mathrm{arcsec}$. We adopted the filters proposed by \cite{Lindegrenetal21} to reduce contamination by binary stars, calibration problems or spurious astrometric solutions. We then followed a similar prescription to the one we had used before \citep{BashiZucker19} to derive the three-dimensional velocity vector $(U, V, W)$ of each star following the procedure described by \citet{JohnsonSoderblom87}. To further reduce contamination of our sample by multiple star systems, bad label determinations and inaccurate elemental abundance derivation, we adopted some of the quality flags suggested by \cite{Xiangetal19}. Thus, we excluded stars having spectral $\mathrm{S/N}_g < 30$, low quality of the spectral fits (\texttt{QFLAG\textunderscore CHI2 = 'bad'}) and binary and multiple star system (\texttt{FLAG\textunderscore SINGLESTAR = 'No'}). 

We limited our analysis to FGK dwarfs with $\mathrm{log}\,g>4$ and $T_\mathrm{eff}=4500-6500~\mathrm{K}$, and excluded metal-poor ($\mathrm{[Fe/H]}$ < $-1$) stars as suggested by \cite{Xiangetal19} in order to avoid the bias caused by the limited training set of metal-poor stars they had used in this region. 

\begin{figure} 
\centering
\includegraphics[width=8.2cm]{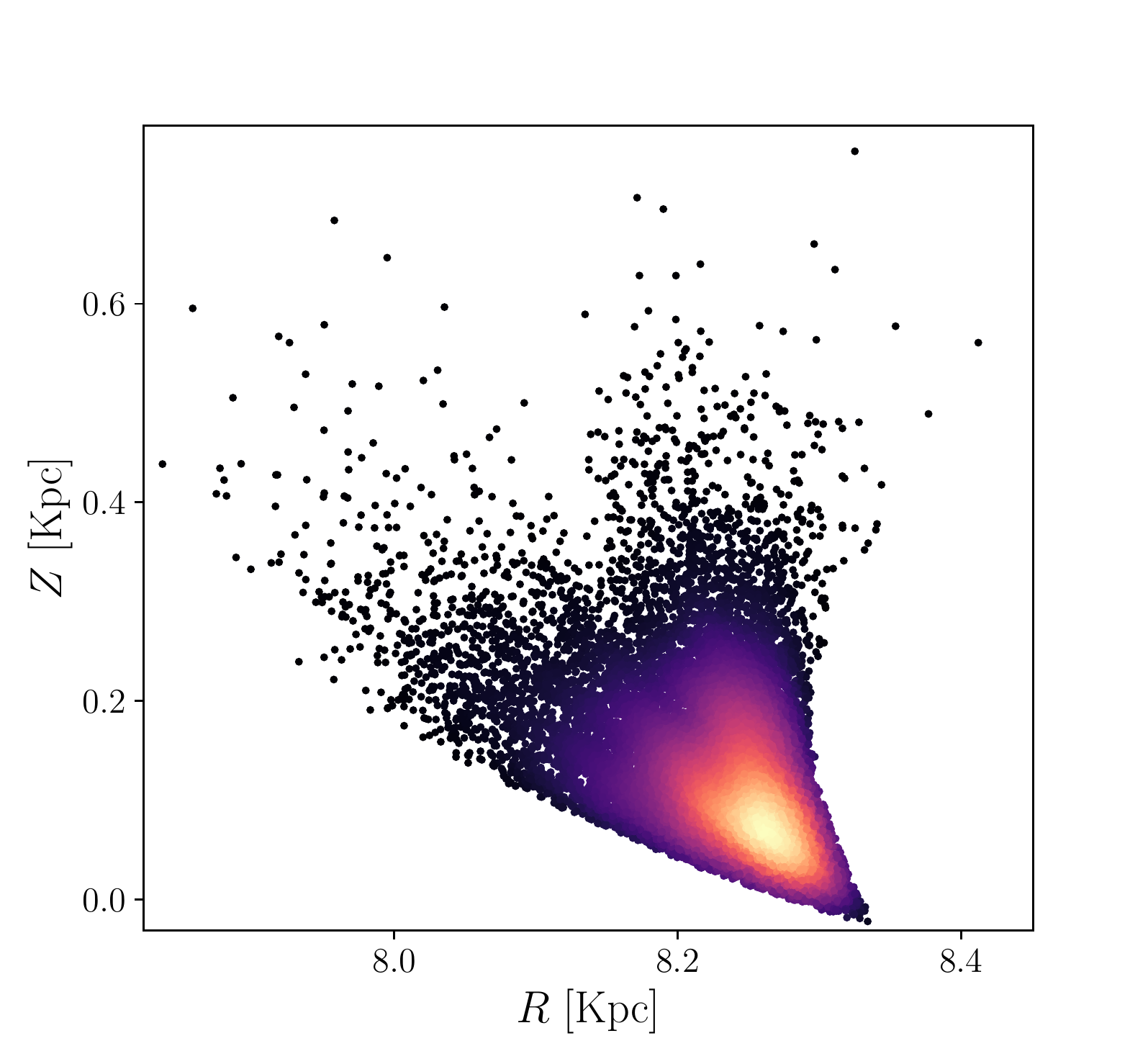}
\includegraphics[width=8.2cm]{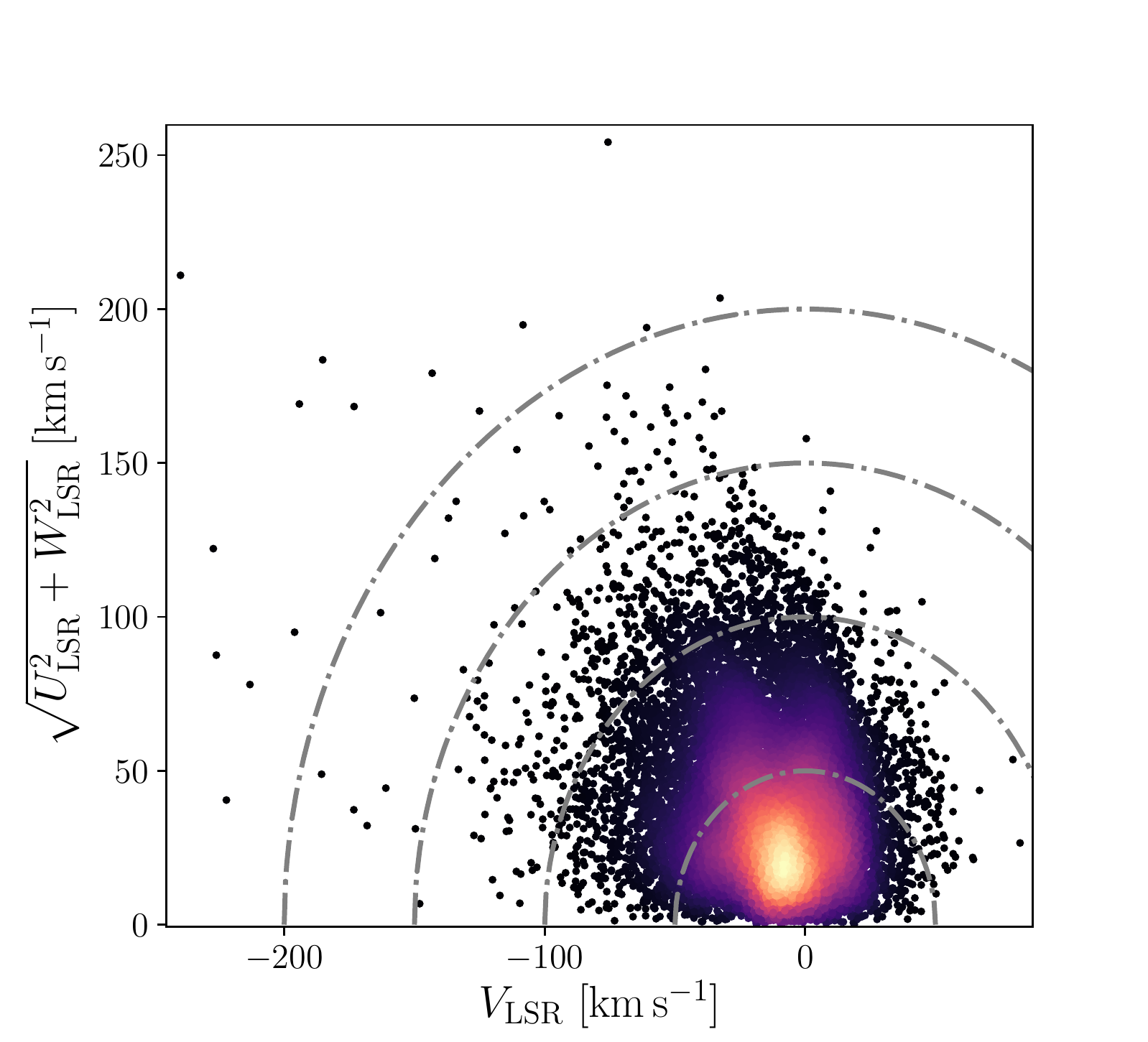}
\includegraphics[width=8.2cm]{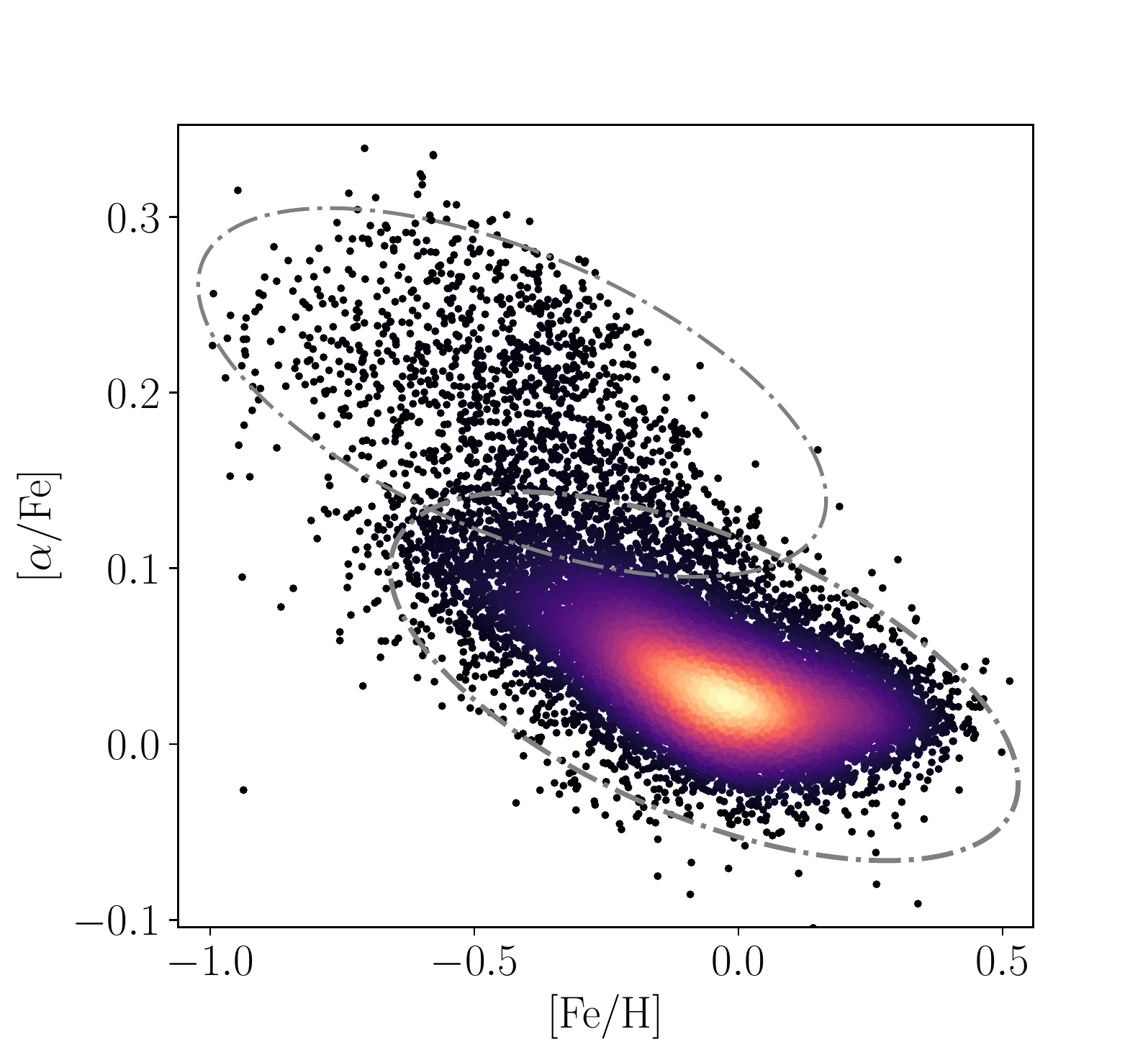}
\caption{\textit{Top:} Position in Galactocentric coordinates of radius ($R$)-height ($Z$) of our sample of $15881$ stars. \textit{Middle:} A Toomre diagram of our sample. \textit{Bottom:} A scatter plot of the same sample in the $\mathrm{[Fe/H]}$- $\mathrm{[\alpha/Fe]}$ plane. The colour represents the density of the points.}
\label{Figure:AllSample}
\end{figure}

After applying these cuts, we were left with a sample of $15881$ KIC stars (Table~\ref{table:KICsample}) with both Galactic space velocities relative to the Local Standard of Rest (LSR; $U_\mathrm{LSR}, V_\mathrm{LSR}, W_\mathrm{LSR}$) as well as $\mathrm{[Fe/H]}$ and $\mathrm{[\alpha/Fe]}$. Figure~\ref{Figure:AllSample} presents the stellar sample in Galactocentric coordinates of radius ($R$)-height ($Z$) (upper panel), using a a Toomre diagram (middle panel) and on the $\mathrm{[Fe/H]}$-$\mathrm{[\alpha/Fe]}$ plane (bottom panel).

\begin{table*}
\caption{List of our final KIC stellar sample including sky position in Equatorial coordinates, parallax, position in Galactocentric coordinates, spectral properties, Galactic space velocities relative to the Local Standard of Rest, relative Galactic component membership, element content and elemental cluster affiliation probabilities. The table is  sorted by KIC number. The full table is available online.}

\label{table:KICsample} 
\centering  
\begin{tabular}{c c c c c c c c c} 
\hline  \hline      

KIC Number & RA [deg] & DEC [deg]  & $\pi$ [mas] & $R$ [pc] & $Z$ [ps] &  $R_\mathrm{*}$  [$R_\mathrm{\odot}$] & $\mathrm{log}g$ [dex] & $T_\mathrm{eff}$ [$\mathrm{K}$]  \\  
\hline                   
$1025494$ & $290.90301$ & $36.76722$ & $2.583\pm0.0107$ & $8301.24$ & $10.50$ & $1.452\pm0.108$ & $4.074\pm0.039$ & $5724\pm18$ \\

$1025986$ & $291.03369$ & $36.77104$ & $9.168\pm0.063$ & $8303.26$ & $0.31$ & $0.979\pm0.041$ & $4.449\pm0.073$ & $5698\pm32$\\

$1026669$ & $291.19248$ & $36.74039$ & $2.566\pm0.009$ & $8304.72$ & $3.55$ & $1.131\pm0.084$ & $4.201\pm0.045$ & $5969\pm21$\\

$1026838$ & $291.22889$ & $36.74170$ & $1.299\pm0.012$ & $8334.08$ & $-22.21$ & $1.086\pm0.082$ & $4.345\pm0.055$ & $5887\pm23$\\

$1026911$ & $291.24386$ & $36.78153$ & $1.743\pm0.001$ & $8299.00$ & $16.89$ & $1.502\pm0.066$ & $4.147\pm0.043$ & $6160\pm20$\\
\hline \hline    
\end{tabular}
\end{table*}
\addtocounter{table}{-1}
\begin{table*}
\caption{continued} 
\label{table:KICsampleGal} 
\centering  
\begin{tabular}{c c c c c c c} 
\hline  \hline      

 $\left(U_\mathrm{LSR},V_\mathrm{LSR},W_\mathrm{LSR} \right)~\mathrm{[km~ s^{-1}]}$  &   $TD/D$ & $TD/H$ & $\mathrm{[Fe/H]}$ [dex] &  $\mathrm{[\alpha/Fe]}$ [dex] & $P_{\mathrm{\alpha}}$  \\
\hline  

($112.46\pm1.41$ , $-53.38\pm3.66$ , $17.97\pm0.68$) & $6.81\pm1.29$ & $1618\pm99$ & $-0.003\pm0.030$ & $0.029\pm0.013$ & $0.9999912\pm1.26\times10^{-5}$\\

($-19.77\pm2.47$ , $-24.79\pm6.53$ , $-3.08\pm1.22$) & $0.0169\pm0.0055$ & $10387\pm903$ & $0.015\pm0.046$ & $-0.002\pm0.026$ & $0.99997\pm2.7\times10^{-4}$\\

($15.82\pm1.80$ , $11.21\pm4.80$ , $29.76\pm0.88$) & $0.0449\pm0.011$ & $8345\pm514$ & $-0.481\pm0.043$ & $0.093\pm0.015$ & $0.989\pm3.16\times10^{-2}$\\

($64.89\pm2.13$ , $13.58\pm5.48$ , $6.26\pm1.00$) & $0.069\pm0.019$ & $12290\pm677$ & $0.0244\pm0.039$ & $0.030\pm0.020$ & $0.99988\pm4.77\times10^{-4}$\\

($4.945\pm2.04$ , $-26.17\pm5.41$ , $-1.77\pm0.99$) & $0.0152\pm0.0035$ & $10642\pm690$ & $-0.173\pm0.033$ & $0.019\pm0.016$ & $0.9999915\pm2.58\times10^{-5}$\\
\hline \hline    
\end{tabular}
\end{table*}

We then went on to classify the stars according to the Galactic populations: 

From a kinematic perspective, we used the probabilistic approach first introduced by \citet{Bensbyetal03} and further implemented in many other works \citep[e.g.][]{Reddyetal06, Bensbyetal14,  Ganetal20,  Chenetal21}. In its essence, we assume the Galactic space velocities ($U_\mathrm{LSR}, V_\mathrm{LSR}, W_\mathrm{LSR}$) of the stellar populations have Gaussian distributions:

   \begin{equation}
      f = k \times \mathrm{exp} \left( 
      -\frac{U_\mathrm{LSR}^2}{2{\sigma_U}^2}
      -\frac{\left( {V_\mathrm{LSR}-V_\mathrm{asym}}\right)^2}{2{\sigma_V}^2}
      -\frac{W_\mathrm{LSR}^2}{2{\sigma_W}^2}
      \right)\, ,
   \end{equation}
where 
   \begin{equation}
      k  = \frac{1}{\left(2 \pi\right)^{3/2}\sigma_U\sigma_V\sigma_W}\, ,
   \end{equation}
{is the normalization coefficient, $\sigma_U$, $\sigma_V$, $\sigma_W$ are the characteristic velocity dispersions and $V_\mathrm{asym}$ is the
asymmetric drift.} 

We adopted the revised kinematic characteristics at different Galactocentric radii ($R$) and heights ($Z$) as described in Table~2 of \cite{Chenetal21}, and calculated the relative Galactic-component membership probabilities of each individual star: $TD/D$ (thick-disk-to-thin-disk), $TD/H$ (thick-disk-to-halo) using: 
   \begin{equation}
      \frac{TD}{D}=\frac{X_\mathrm{TD}}{X_\mathrm{D}}\cdot \frac{f_\mathrm{TD}}{f_\mathrm{D}},~~ \frac{TD}{H}=\frac{X_\mathrm{TD}}{X_\mathrm{H}}\cdot \frac{f_\mathrm{TD}}{f_\mathrm{H}}\, ,
   \end{equation}
where $X$ is a variable representing the fraction of stars for a given galactic component and depends on the stellar position in the Galaxy.

From elemental abundance perspective, following \cite{Blancatoetal19} and \cite{Bashietal20}, we used a Gaussian Mixture Model (GMM) classifier and fixed the number of clusters to two in order to capture the bimodality in the $\mathrm{[Fe/H]}$-$\mathrm{[\alpha/Fe]}$ plane. Consequently, each point is affiliated with one of the two clusters with some probability $P_\alpha$, where values closer to $1$ suggest higher probability to belong to the solar-$\alpha$ cluster while values closer to $0$ suggest higher probability to belong to the $\alpha$-enriched cluster. It is important to note that as we consider uncertainties in the stellar elemental content in our final planet occurrence rate estimates, \cite[as opposed to][]{Bashietal20}, $P_{\mathrm{\alpha}}$ is also accompanied by an uncertainty estimate. 

Thus, we defined our sample of stars in the Galactic context as (i) thin disk stars only if they where found to be within our GMM solar-$\alpha$ cluster with $P_{\mathrm{\alpha}} > 0.8$ and $TD/D < 0.1$; (ii) thick disk stars if they where found to be within the $\alpha$-enriched cluster with $P_{\mathrm{\alpha}} < 0.2$, $TD/D > 2$ and $TD/H > 1$ and (iii) halo stars, if , $TD/D > 10$ and $TD/H < 1$ (ignoring the elemental content).

\begin{figure*}
\centering
\includegraphics[width=12cm, height=12cm] {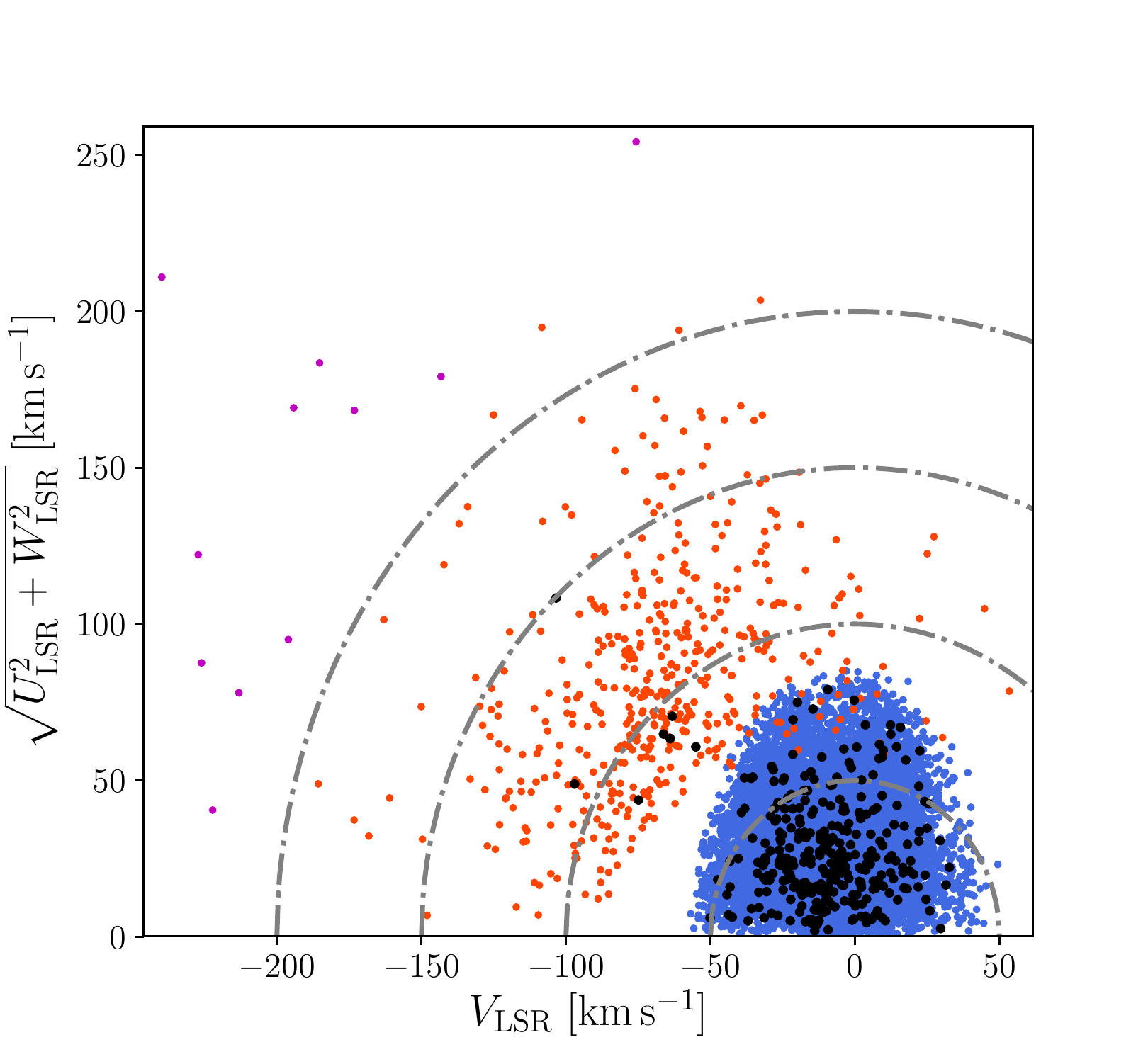}
\caption{Toomre diagram of the parent stellar sample (blue dots: thin disk, red dots: thick disk, purple dots: halo stars) and planet-host stellar sample (black dots) of our Galactic component samples (after applying cuts on the kinematics and element abundance as detailed in the main text). Dotted lines indicate constant total velocities $V_{\mathrm{tot}}$ at $50$, $100$, $150$ and $200  \mathrm{km~ s^{-1}}$. }
\label{Figure:FigGalacticSample}
\end{figure*}

\begin{figure*}
\centering
\includegraphics[width=8.0cm]{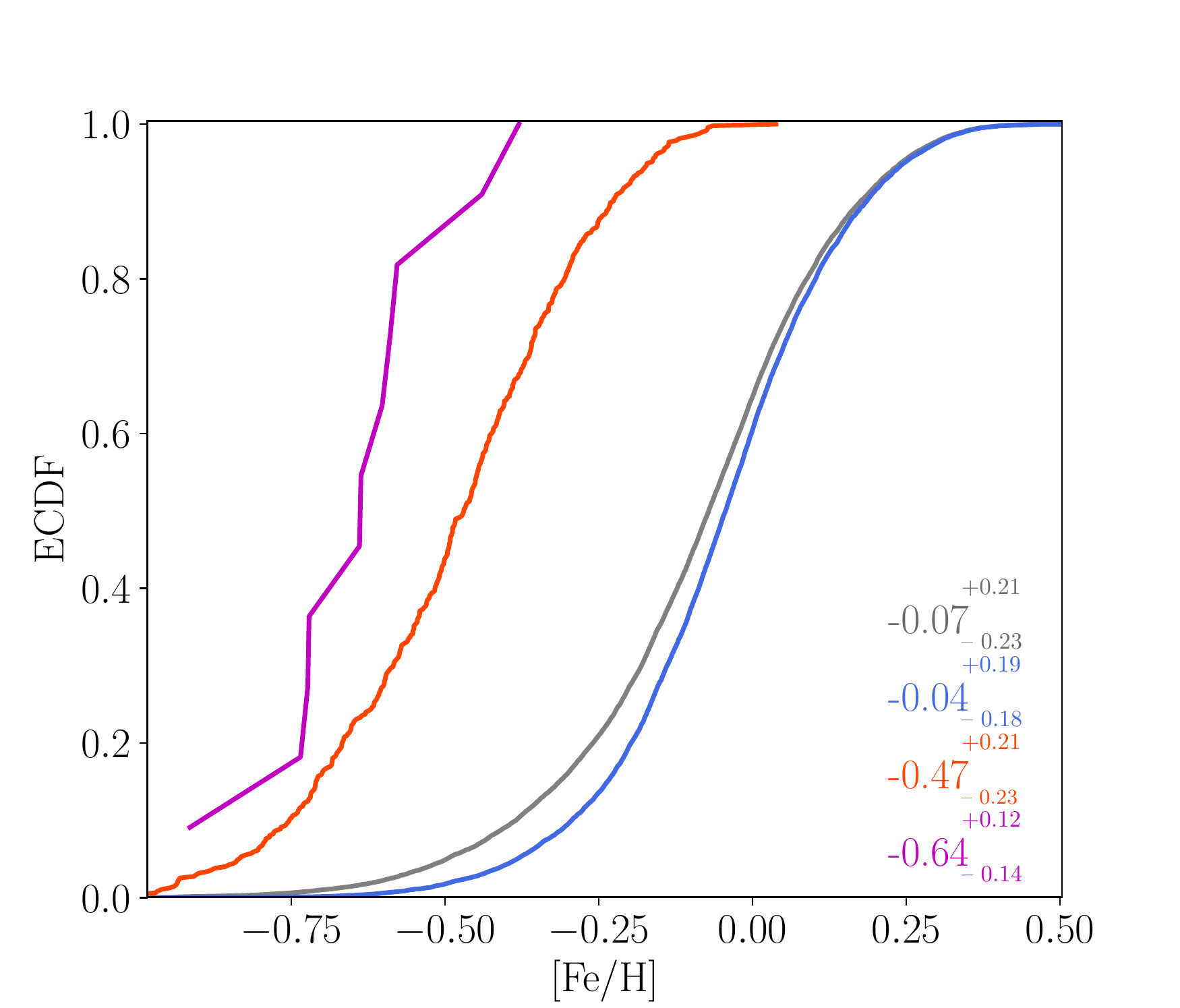}
\includegraphics[width=8.0cm]{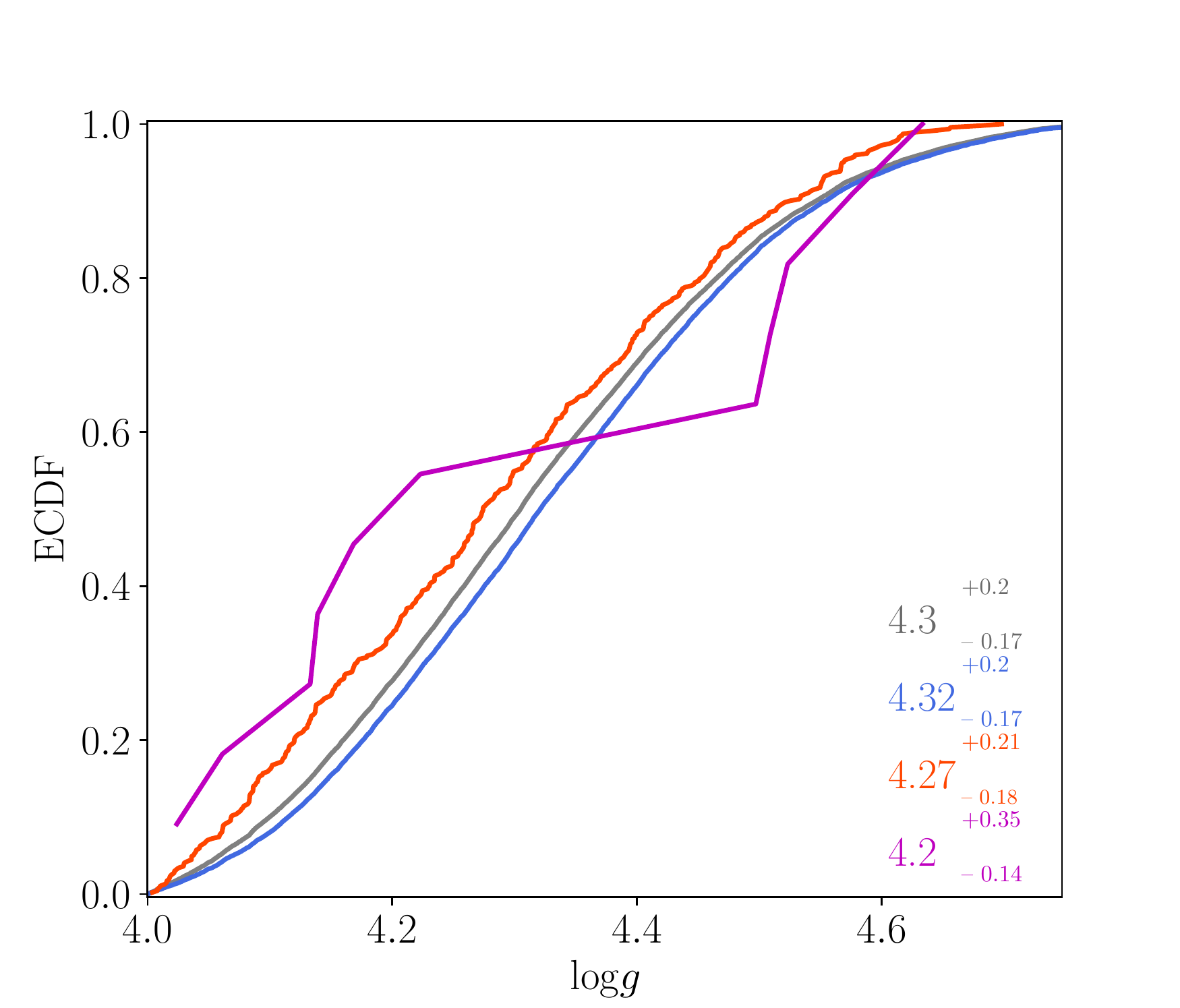}
\includegraphics[width=8.0cm]{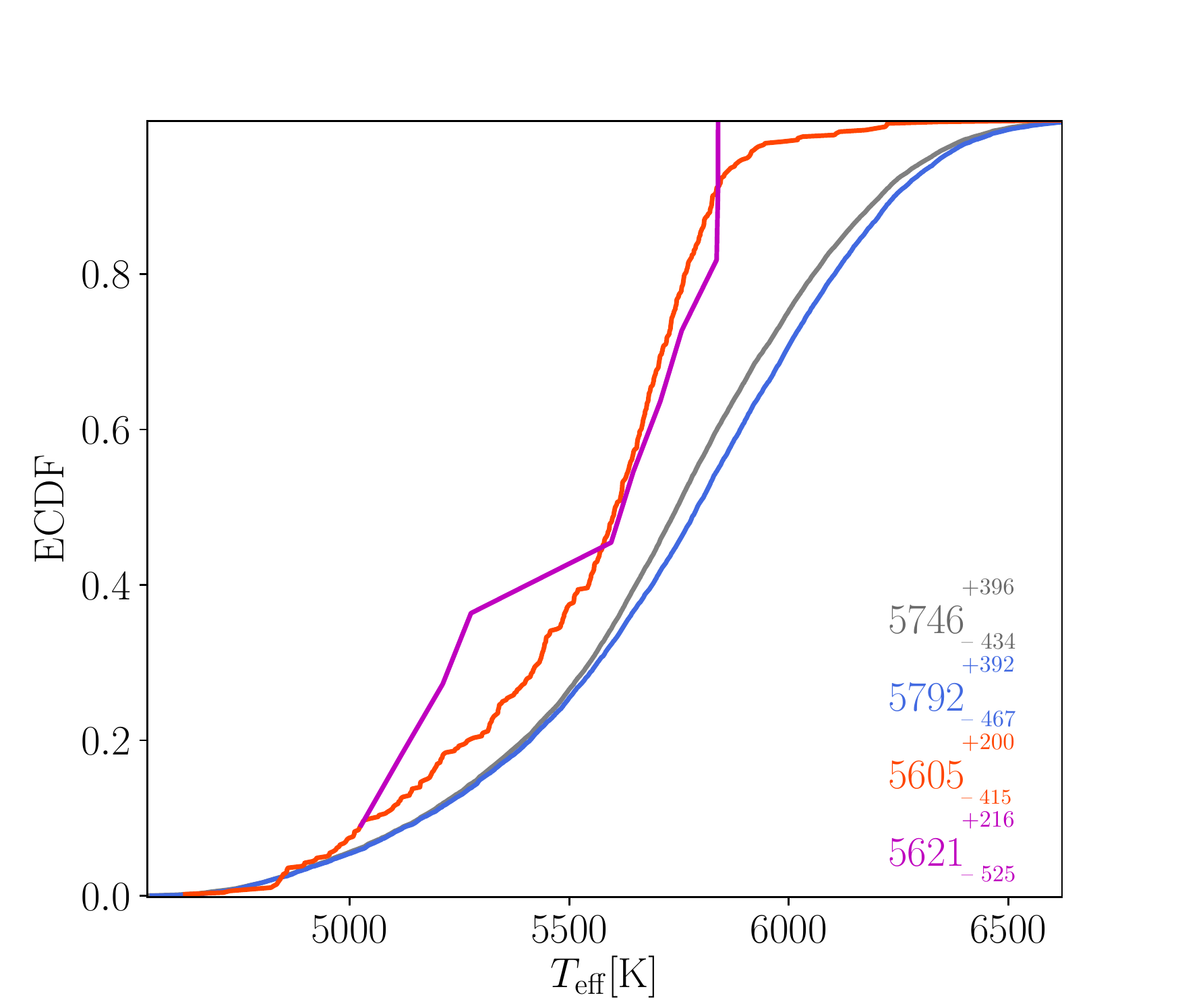}
\includegraphics[width=8.0cm]{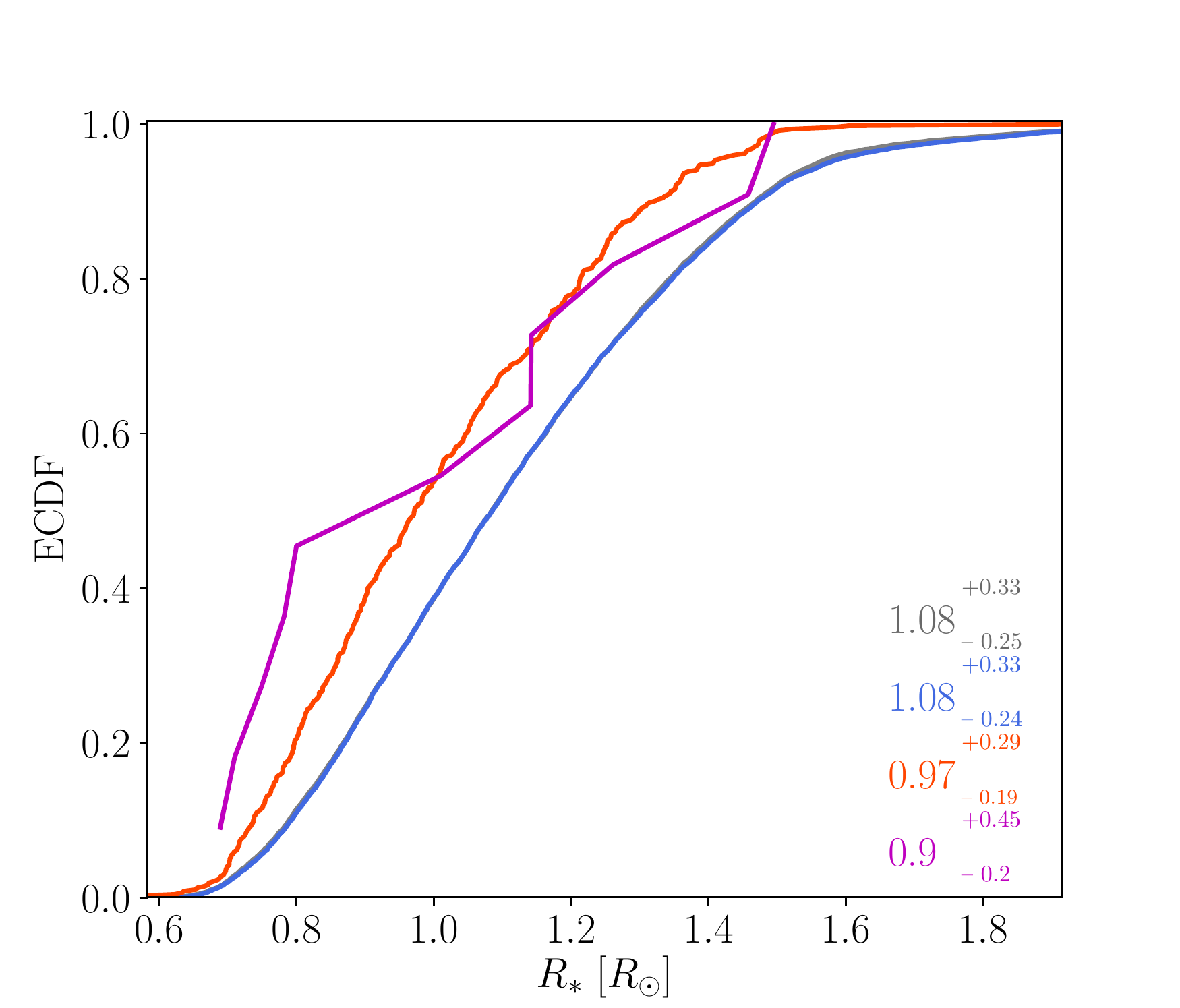}

\caption{Empirical cumulative distribution function (ECDF) and $16$th, $50$th, $84$th percentiles (bottom right side of each panel) of the stellar properties ($\mathrm{[Fe/H]}$, $\mathrm{log}g$, $T_\mathrm{eff}$, $R_\mathrm{*}$) of our different Galactic population samples: parent sample (gray), thin disk (blue), thick dist (red) and halo (purple).}
\label{ECDFGalactic}
\end{figure*}

Our final stellar sample consisted of $11401$ thin-disk stars, $472$ thick-disk stars and $11$ halo stars. As can be clearly seen in the Toomre diagram of our final stellar samples (Figure \ref{Figure:FigGalacticSample}), stars with relatively low total velocities $V_{\mathrm{tot}} = \sqrt{U_\mathrm{LSR}^2+V_\mathrm{LSR}^2+W_\mathrm{LSR}^2} < 50\,\mathrm{km~ s^{-1}}$ have been conclusively assigned to the thin disk, while higher total velocities $V_{\mathrm{tot}}=90-200\,\mathrm{km~ s^{-1}}$ and $V_{\mathrm{tot}} > 200\,\mathrm{km~ s^{-1}}$ were associated with the thick disk and stellar halo respectively \citep{Bensbyetal03, NissenSchuster10, Bensbyetal14, BashiZucker19, Chenetal21}. We compare in Figure~\ref{ECDFGalactic} the distributions of the stellar properties of iron content ($\mathrm{[Fe/H]}$), gravity ($\log g$), effective temperature ($T_\mathrm{eff}$) and radius ($R_\mathrm{*}$) among our Galactic component samples. As expected, the thin-disk solar-$\alpha$ cluster of metal-rich stars in our sample, appears to be hotter and larger, compared to the thick disk and halo star samples. Using a Kolmogorov–Smirnov (KS) test, we obtain extremely low p-values suggesting clear differences in the distributions of stellar properties among the samples of the Galactic components. For example, the $\mathrm{\log_{10}}$(p-value) of the KS-test comparison between the $\mathrm{[Fe/H]}$ distributions of the thin and thick disk samples is $-217$. 

\subsection{Planet sample} 
\label{Planetsample}
We based our planet sample on the $Kepler$ objects of interest (KOI) catalogue from the $25$th data release (DR25) of \textit{Kepler} \citep{Thompsonetal18}. We kept only KOIs that were associated with our parent stellar catalogue, and that were designated as planet candidates with a disposition score higher than $0.9$ (similar thresholds were also used by \citet{Muldersetal18} and \citet{Yangetal20}). Out of this sample, we excluded objects flagged as false positives. Using updated stellar radii from \textit{Gaia} DR2 \citep{GaiaCol18} and the transit depths listed in \textit{Kepler} DR25, we revised the KOI planet radii. Furthermore, we excluded from our sample grazing transiting planets with $b > 1 - R_{\mathrm{p}}/R_{\mathrm{*}} $ as listed in the KOI catalogue, and also limited our sample to include only planet candidates with periods $P < 400$ days and radii in the range $R_\mathrm{p} = 0.5-12\,R_\oplus$.

In total, our sample included $506$ planet candidates orbiting $369$ planet-host stars, out of which $402$ planets and $292$ host stars included in the thin-disk sample, eight planets and seven host stars in the thick-disk sample and no detected planets in our stellar halo sample. Table~\ref{table:KOIsample} details the full list of KOIs used in this work, and Figure~\ref{Figure:FigGalacticSample} presents a Toomre diagram of the parent stellar sample (gray) together with the sample of planet-host stars (orange) 

\begin{table}
\caption{List of our final KOI planet sample sorted by KOI number. The full table is available online.} 
\label{table:KOIsample} 
\centering  
\begin{tabular}{c c c} 
\hline  \hline      

 KOI & Orbital Period [day] &  $R_{\mathrm{p}}$  [$R_\oplus$]   \\    

\hline                   
K00049.01 & $8.31377304 \pm 0.00004206$ & $3.773 \pm 0.247$  \\
K00063.01 & $9.43414171 \pm 0.00002136$ & $5.769 \pm 1.246$  \\
K00069.01 & $4.72674037 \pm 0.00000157$ & $1.643 \pm 0.105$  \\
K00072.02 & $45.29422297 \pm 0.000056$ &  $2.354 \pm 0.135$  \\
K00082.01 & $16.14567193 \pm 0.00000744$ & $2.431 \pm 0.202$  \\
\hline \hline    
\end{tabular}
\end{table}

\section{Occurrence-rate Methodology} \label{Occurrence}
Our main aim in this work was comparing the occurrence rates of the Galactic stellar populations. We preferred to avoid any physical model assumptions and infer the occurrence rates based on the data alone. 
We chose to apply a simplified Bayesian model (SBM) framework \citep{Hsuetal18, Bashietal20} using conjugate priors to estimate the two basic planet occurrence rates: (i) the average number of planets per star, $\bar{n}_\mathrm{p}$, and (ii) the fraction of stars with planets, $F_\mathrm{h}$.

In this model, the posterior distribution of $\fpl$ would be a gamma distribution with shape parameter $\alpha_0 + \Ntotpl$ and rate parameter $\beta_0 + \Neffn$:
   \begin{equation}
      p(\fpl | \Ntotpl, \Neffn) \sim \mathrm{Gamma}(\alpha_0 + \Ntotpl, \beta_0 + \Neffn) \, ,
   \end{equation}
where $\alpha_0$ and $\beta_0$ are the shape and rate parameters of the prior, $\Ntotpl$ is the number of planets detected around stars in the sample, and $\Neffn$ is the effective number of stars searched for the detection of planets (see Appendix \ref{Appendix:EffectiveStars} for more details). We assume an uninformative prior with $\alpha_0 = \beta_0 = 0$.\footnote{This is equivalent to a log-uniform distribution over all the real numbers, which is an 'improper prior', but is nevertheless justified for use in certain Bayesian contexts \citep[e.g.][]{TarLin2010}.}

Similarly, the posterior distribution for $\FplH$ is a beta distribution with shape parameters $\tilde{\alpha_0} + \NtotplHost$ and $\tilde{\beta_0} + \NeffF-\NtotplHost$:
   \begin{equation}
      p(\FplH | \NtotplHost, \NeffF) \sim \mathrm{Beta}(\tilde{\alpha_0} + \NtotplHost, \tilde{\beta_0} + \NeffF-\NtotplHost) \,,
   \end{equation}
where $\tilde{\alpha_0}$ and $\tilde{\beta_0}$ are the shape and rate parameters of the prior, $\NtotplHost$ is the number of planet-host stars in a sample, and $\NeffF$ is the effective number of stars searched for the detection of planetary systems (see Appendix \ref{Appendix:EffectiveStars} for more details). The least restrictive prior for $\FplH$ is obviously a uniform distribution between $0$ and $1$, which is equivalent to a beta distribution with $\tilde{\alpha} = \tilde{\beta} = 1$.

Note that, in general, the effective number of stars searched is estimated based on the probability of detection, which is different in the cases of individual planets or planetary systems, and therefore $ \Neffn \neq \NeffF$.

In order to account for the uncertainties in stellar properties, the values of planet occurrence rates (both $\bar{n}_\mathrm{p}$ and $F_\mathrm{h}$) that we report in the next section are based on the following recipe: we resampled the stellar sample by drawing the stellar properties of each star from Gaussian distributions centered around the published nominal values, with standard deviations determined by the published errors. We repeated this resampling $100$ times, each time estimating again $\bar{n}_\mathrm{p}$ and $F_\mathrm{h}$. For each one of those $100$ resamplings, we generated $1000$ random draws from the derived posterior distributions of $\bar{n}_\mathrm{p}$ and $F_\mathrm{h}$. The final values we report are the median and $16-84\%$ ($1\sigma$) confidence intervals of the $100 \times 1000$ cases, as the posterior planet occurrence rate estimates and uncertainties.


\section{Results}
\label{Results}
By now quite a few occurrence-rate studies have been published that are based on \textit{Kepler} stars, using various methodologies and sample definitions \citep{Fultonetal17, Zhuetal18, Muldersetal19, Hsuetal19, Heetal19, Yangetal20, KunimotoMatthews20}. We therefore decided first to perform a sanity check of our analysis by comparing our estimates with previous studies, ignoring the Galactic context. Reassuringly, we found our results to be consistent with these works: 

(i) We found the average number of planets per star ($\fpl$) at the parameter range of $P=1-100$\,days, $R_\mathrm{p} = 1-2\,R_\oplus$ to be $0.399^{+0.029}_{-0.028}$, in agreement with the estimates of $\sim 0.40 \pm 0.5$\footnote{Value deduced from Figure~2. of \cite{Hsuetal19}} and $0.419 \pm 0.012$ of \cite{Hsuetal19} and \cite{Muldersetal19} respectively for similar parameter ranges. At the $P=1-100$\,days and $R_\mathrm{p}=2-4\,R_\oplus$ range, we obtained a value of $0.418^{+0.033}_{-0.031}$ that is comparable to the estimates by \cite{Fultonetal17} and \cite{KunimotoMatthews20} of $0.37\pm 0.02$ and $0.35\pm 0.02$ respectively. 

(ii) As for the fraction of stars with planets ($\FplH$), the value we found at the $P=3-300$\,days and $R_\mathrm{p}=0.5-10\,R_\oplus$ region was $0.441\pm0.019$, while previous works that examined similar domains of planet properties suggested a wider range of occurrence rates ranging from $0.30 \pm 0.03$ \citep{Zhuetal18} to $0.52^{+0.04}_{-0.03}$ \citep{Yangetal20} and $0.56^{+0.04}_{-0.03}$ \citep{Heetal20}.

We list in Table \ref{table:OccResults} our estimated planet occurrence rates in the general sample for representative ranges of planet properties.

\begin{table}
\caption{Planet occurrence rates: average number of planets per star ($\fpl$), and the fraction of stars with planets ($\FplH$) in the general sample}    
\label{table:OccResults}  
\centering             
\begin{tabular}{c c c c}    
\hline  \hline      
$R_{\mathrm{p}}~[R_\oplus]$ & $P~\mathrm{[days]}$ & $\fpl$ & $\FplH$\\    
\hline                   
$1-2$ & $1-100$ & $0.399^{+0.029}_{-0.028}$ & $0.174^{+0.013}_{-0.0125}$ \\
   
$2-4$ & $1-100$ & $0.418^{+0.033}_{-0.031}$ & $0.203\pm 0.013$ \\
   
$1-4$ & $1-100$ & $0.821^{+0.043}_{-0.042}$ & $0.283^{+0.015}_{-0.014}$ \\
   
$4-12$ & $1-100$ & $0.0317^{+0.0101}_{-0.0083}$ & $0.0343^{+0.0103}_{-0.00859}$ \\
   
$0.5-10$ & $3-300$ & $1.694^{+0.083}_{-0.081}$ & $0.441\pm 0.019$ \\
   
\hline \hline                               
\end{tabular}
\end{table}

\subsection{Occurrence rates in the Galactic context}

\begin{figure*}
\centering
\includegraphics[width=8.7cm]{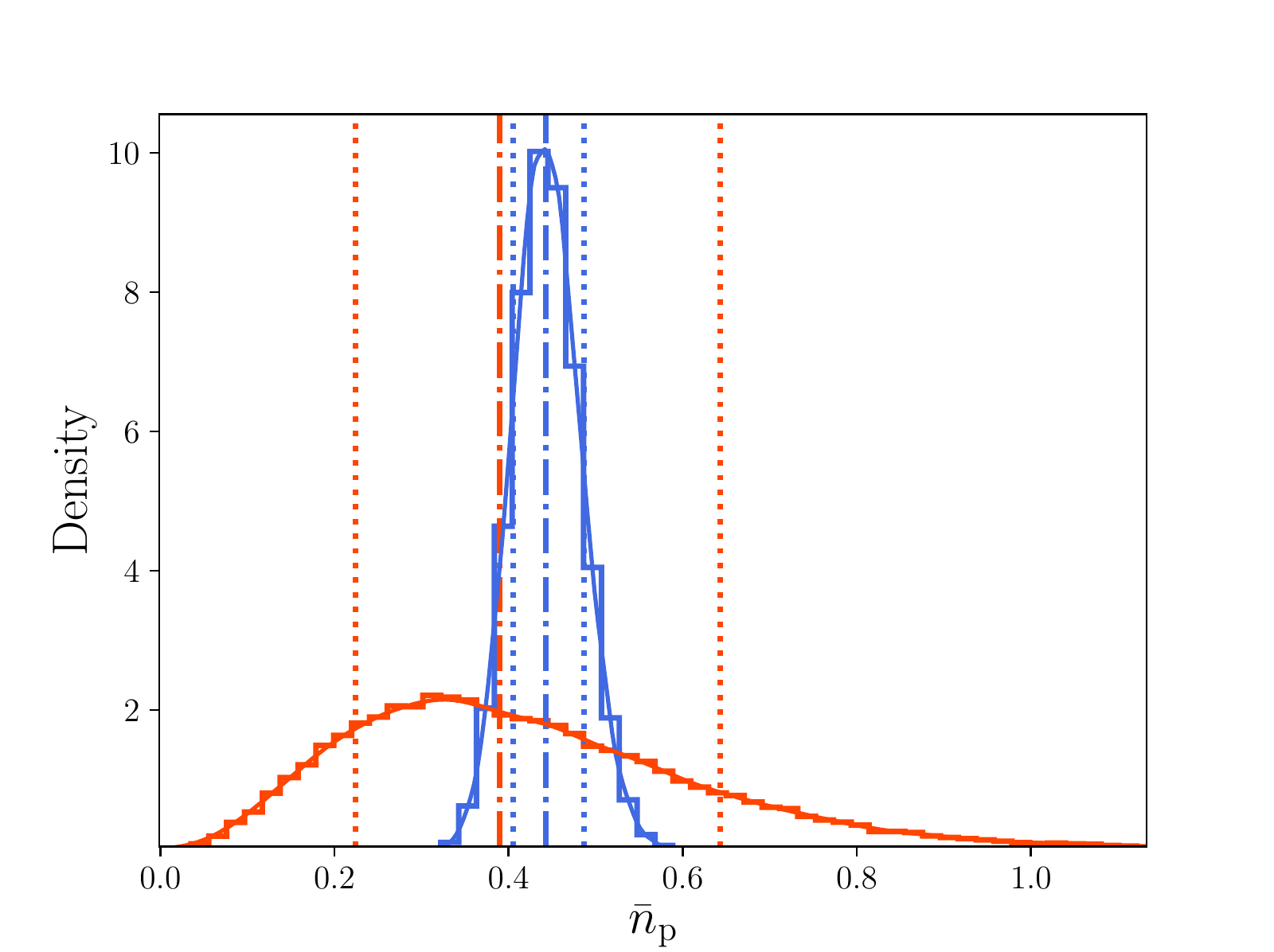}
\includegraphics[width=8.7cm]{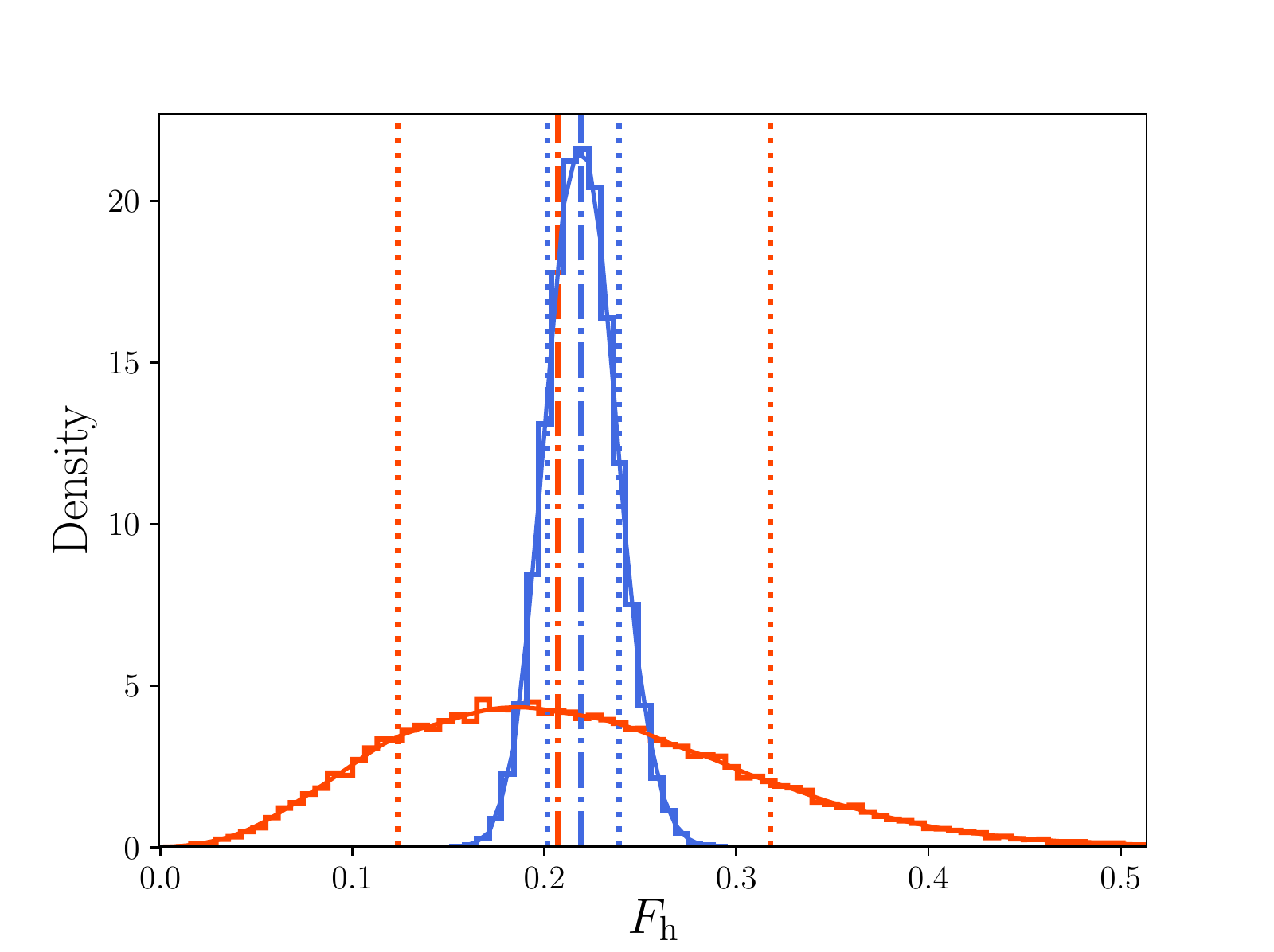}
\caption{Posterior distributions of estimated planet occurrence rates. up: average number of planets per star ($\fpl$), bottom: the fraction of stars with planets ($\FplH$) in the Galactic context of the thin disk: blue histogram, thick disk: red histogram, at the $R_{\mathrm{p}} = 2-4\,R_\oplus$,  $P = 1-100$ days. Vertical lines represents the $16$th, $50$th, $84$th percentiles. The accompanying curves of same colour are a best fit kernel estimations of the posterior distributions.}
\label{Fig.OccEx_1_100_2_4}
\end{figure*}

\begin{figure*}
\centering
\includegraphics[width=8.7cm]{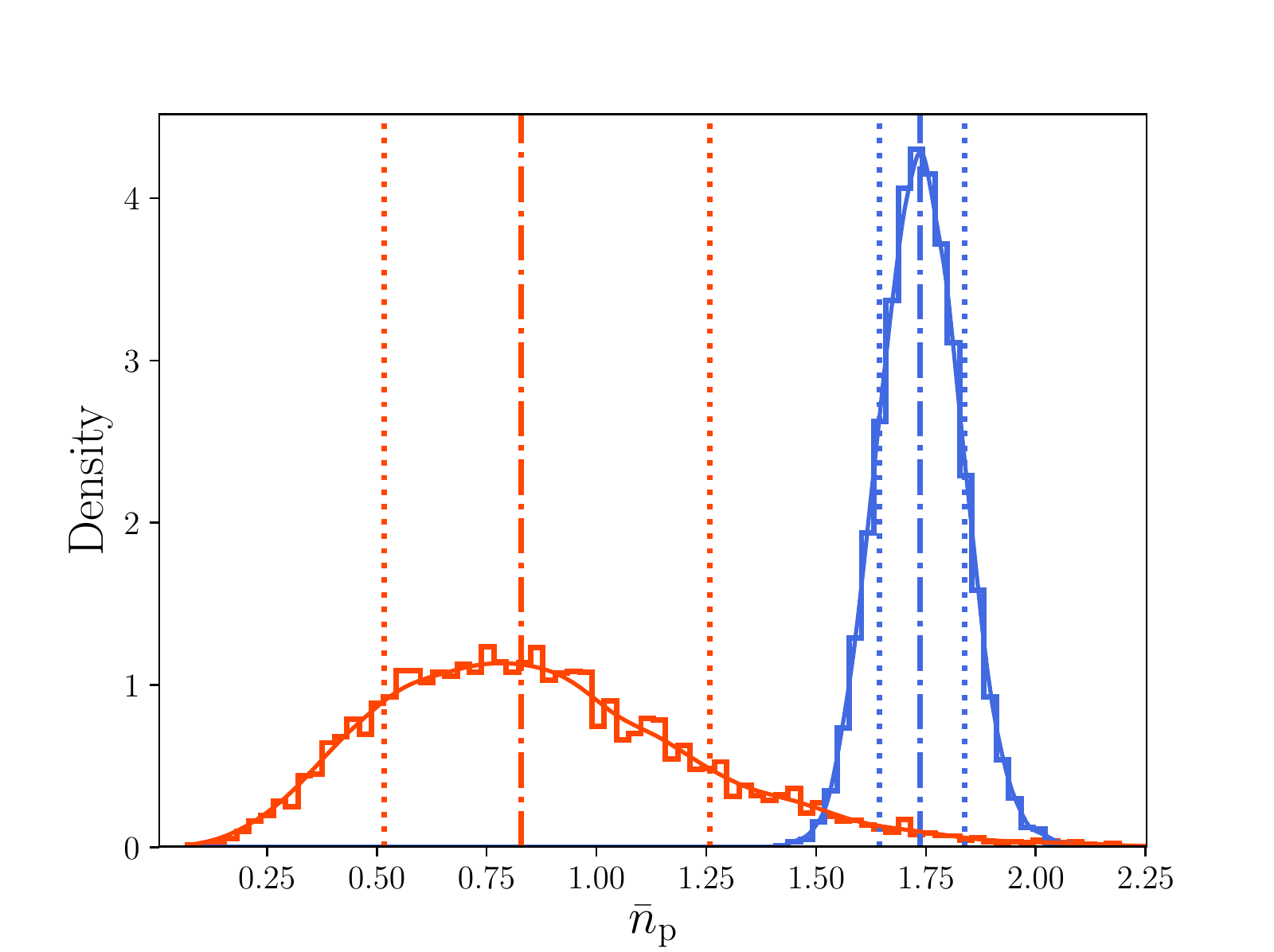}
\includegraphics[width=8.7cm]{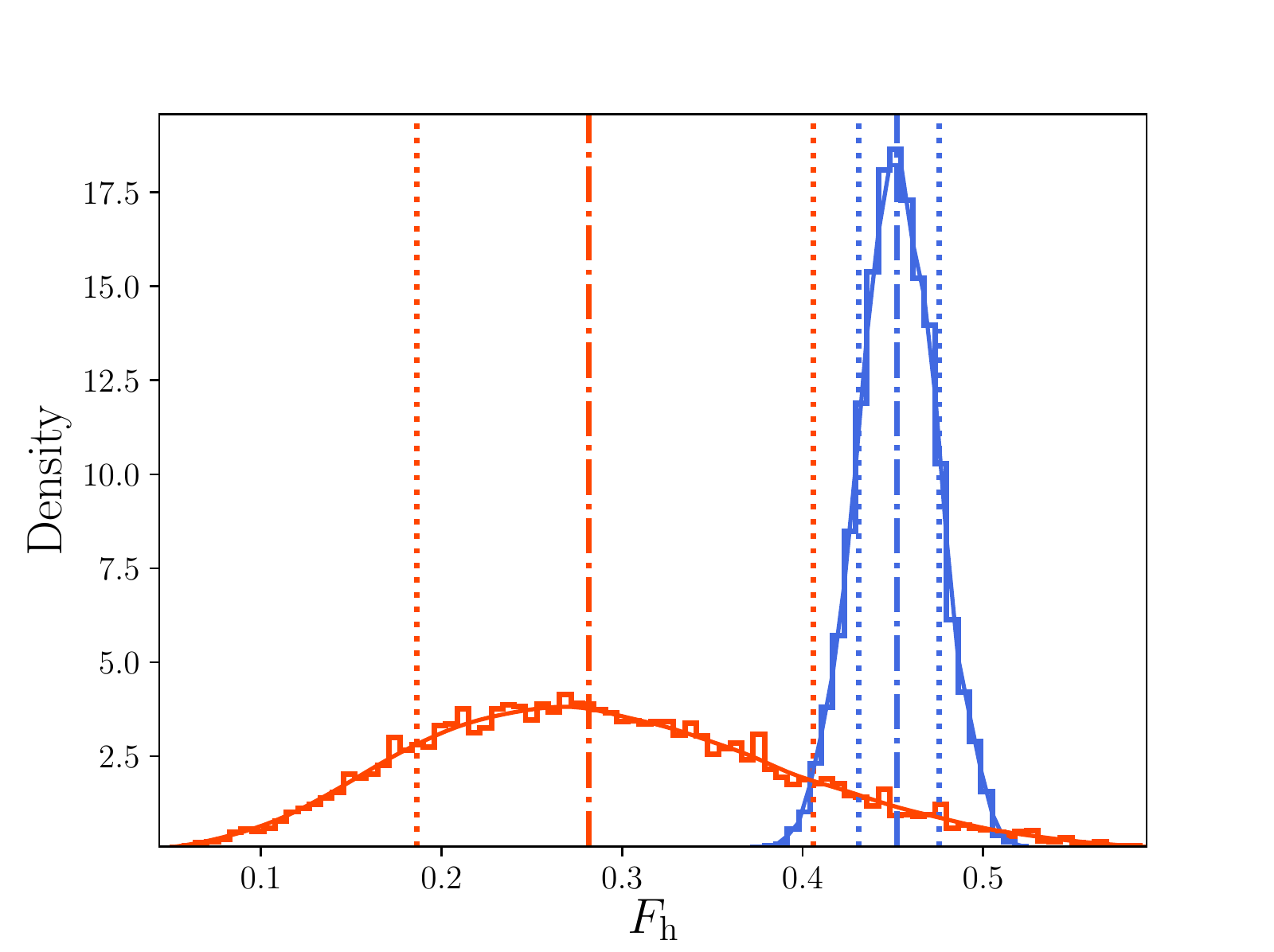}
\caption{Same as Fig. \ref{Fig.OccEx_1_100_2_4} but for $R_{\mathrm{p}} = 0.5-10\,R_\oplus$,  $P = 3-300$ days.}

\label{Fig.OccEx_3_300_0.5_10}
\end{figure*}

Table~\ref{table:GalacticOccResults} lists the estimated planet occurrence rates we obtained for the same representative ranges of planet properties as in Table~\ref{table:OccResults}, after dividing the sample to the Galactic population samples.
As an example for the Bayesian estimates we obtain, Figures~\ref{Fig.OccEx_1_100_2_4} and ~\ref{Fig.OccEx_3_300_0.5_10} present the posterior distributions we obtain for the specific range of planetary properties $R_{\mathrm{p}} = 2-4 \,R_\oplus$, $P = 1-100$~days and $R_{\mathrm{p}} = 0.5-10 \,R_\oplus$, $P = 3-300$~days, respectively. 

Unfortunately, the low fraction of halo stars that we identified in our sample has prevented us from estimating a useful upper limit on the occurrence rates in the halo population. Thus, in what follows we limit our analysis to the two disk samples.

In order to estimate the extent to which the occurrence rate posterior distributions of the Galactic context samples actually represent two different values, we performed a t-test (fifth column in Table~\ref{table:GalacticOccResults}, in which we calculated the mean difference between the two distributions, and normalized it by the standard deviation, i.e., the two standard deviations added in quadrature. This value essentially measures how significantly is the difference between the two values differ from zero.

In general, we find that the planet occurrence rates (both in $\fpl$ and $\FplH$) at the region of close-in super Earths $R_{\mathrm{p}} = 1-2\,R_\oplus$, $P = 1-100$ days are higher among thin-disk stars. Nonetheless, for sub-Neptune planets and above ($R_{\mathrm{p}} > 2\,R_\oplus$), these differences seem to disappear. This is in agreement with our recent work \citep{Bashietal20} that was based on RV analysis in the Galactic context of a HARPS stellar sample. We have argued there that the occurrence rates of close-in Super Earths and Neptune size planets ($R_{\mathrm{p}} = 2-4\,R_\oplus$,  $P = 1-100$ days) is similar in thin and thick disk stars.

\begin{table*}
\caption{Planet occurrence rates: average number of planets per star ($\fpl$), the fraction of stars with planets ($\FplH$) and t-test statistics to quantify the differences between the posterior distributions of the two Galactic populations.}    
\label{table:GalacticOccResults}  
\centering             
\begin{tabular}{c c c c c}    
\hline  \hline      
$R_{\mathrm{p}}~[R_\oplus]$ & $P~\mathrm{[days]}$ & Thin disk ($\fpl$, $\FplH$) & Thick disk ($\fpl$, $\FplH$)& t ($\fpl$, $\FplH$) \\    
\hline                   
   $1-2$ & $1-100$ & $0.438^{+0.039}_{-0.035}$ ;  $0.186^{+0.017}_{-0.015}$  &
   $0.154^{+0.157}_{-0.092}$ ;  $0.116^{+0.086}_{-0.058}$ & $1.75$ ; $0.81$\\
   
   $2-4$ & $1-100$ & $0.443^{+0.043}_{-0.038}$ ;  $0.219^{+0.019}_{-0.018}$  &
   $0.389^{+0.253}_{-0.165}$ ;  $0.207^{+0.111}_{-0.083}$ & $0.21$ ; $0.11$\\
   
   $1-4$ & $1-100$ & $0.877^{+0.057}_{-0.052}$ ;  $0.302^{+0.019}_{-0.017}$  &
   
   $0.52^{+0.27}_{-0.19}$ ;  $0.229^{+0.102}_{-0.078}$ &
   
   $1.28$ ; $0.72$\\
   
   $4-12$ & $1-100$ & $0.022^{+0.016}_{-0.011}$ ;  $0.026^{+0.014}_{-0.011}$ &
   --- ;  $0.052^{+0.091}_{-0.038}$ & --- ; $0.28$\\
   
   $0.5-10$ & $3-300$ & $1.73^{+0.12}_{-0.11}$ ;  $0.462 \pm 0.023$ &
   $0.83^{+0.43}_{-0.31}$ ;  $0.281^{+0.124}_{-0.095}$ & $2.09$ ; $1.39$\\
   
\hline \hline                               
\end{tabular}
\end{table*}

\subsection{Dependence on stellar properties and dynamical history}
\label{DependStelProp}
As was shown in previous works, it seems that stellar spectral type and consequently both stellar mass and effective temperature are related to planet occurrence rates. Specifically, there is a trend of increasing planet occurrence rate (especially of small planets with radii smaller than $2.5\,R_\oplus$) towards late type (smaller and cooler) stars \citep{Muldersetal15, Heetal20, KunimotoMatthews20, Yangetal20}. On the other hand, metallicity and mainly stellar iron content [Fe/H] correlate with planet occurrence rates, especially for large planets with radii larger than $1.5\,R_\oplus$ \citep{Petiguraetal18, Zhu19, Luetal2020, Bashietal20}. 

In order to gain more insight into the planet occurrence rates in the Galactic context, one needs to consider not only these stellar observable properties but also the effect of the dynamical history of the stars in the Galaxy. 

As shown in Section~\ref{Thesample} above, we separated our samples of thin and thick disk stars mainly according to the total velocity (stellar kinematics). Thus, any differences in planet occurrence rates between the samples may be related to some role that the motion and position of stars in the Galaxy might play \citep{BashiZucker19, Daietal21}.
Furthermore, as $\alpha$-enhanced nuclides found in thick disk stars suggest early formation in the history of the Galaxy, stellar age may also be an important factor affecting planet occurrence rates. 

\begin{figure}
\centering
\includegraphics[width=8.7cm]{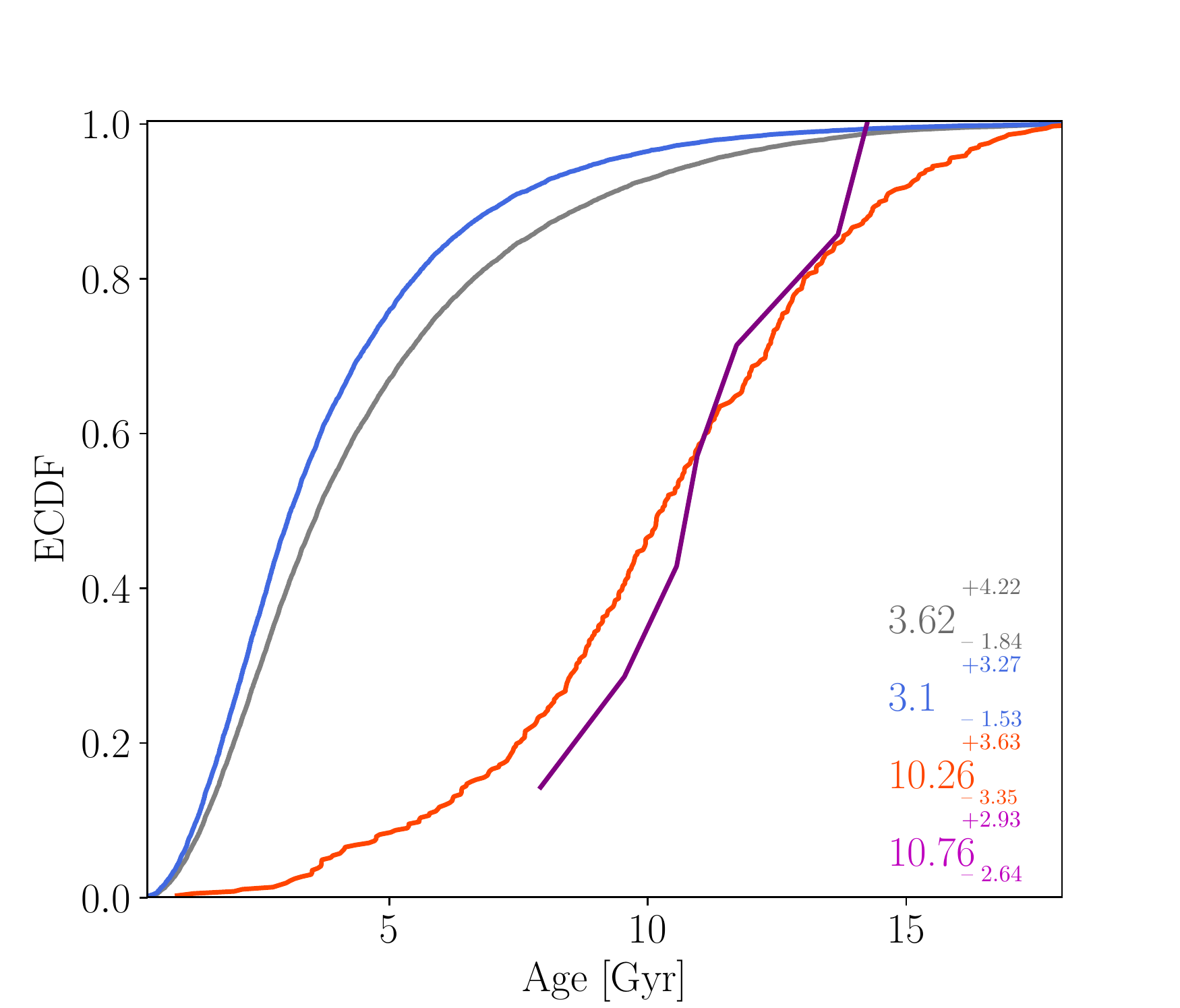}
\caption{Age ECDF, with $16$th, $50$th, and $84$th percentiles (bottom right side of each figure) for our different stellar samples: general sample (gray), thin disk (red), thick disk(blue), and halo (purple).}
\label{Figure:ECDFAge}
\end{figure}

To further investigate those potential relations, we cross-matched our sample with the catalog of \cite{Bergeretal20} who used isochrones to derive stellar age estimates for \textit{Kepler} stars. As can be seen in Figure \ref{Figure:ECDFAge}, a clear distinction in age is apparent, where thin disk stars are much younger than thick disk and halo stars. This dependence persists when we take into account the large uncertainties usually accompanying stellar age estimates.  

To better characterise these effects, we first checked whether we could see the same trends as those suggested in previous works, i.e. the dependence of planet occurrence rates on spectral type ($T_\mathrm{eff}$) and metallicity ([Fe/H]). However, we also tested for dependence on kinematics ($V_{\mathrm{tot}}$) and stellar age, which were not addressed in those previous works. 

\begin{figure*}
\centering
\includegraphics[width=8.3cm]{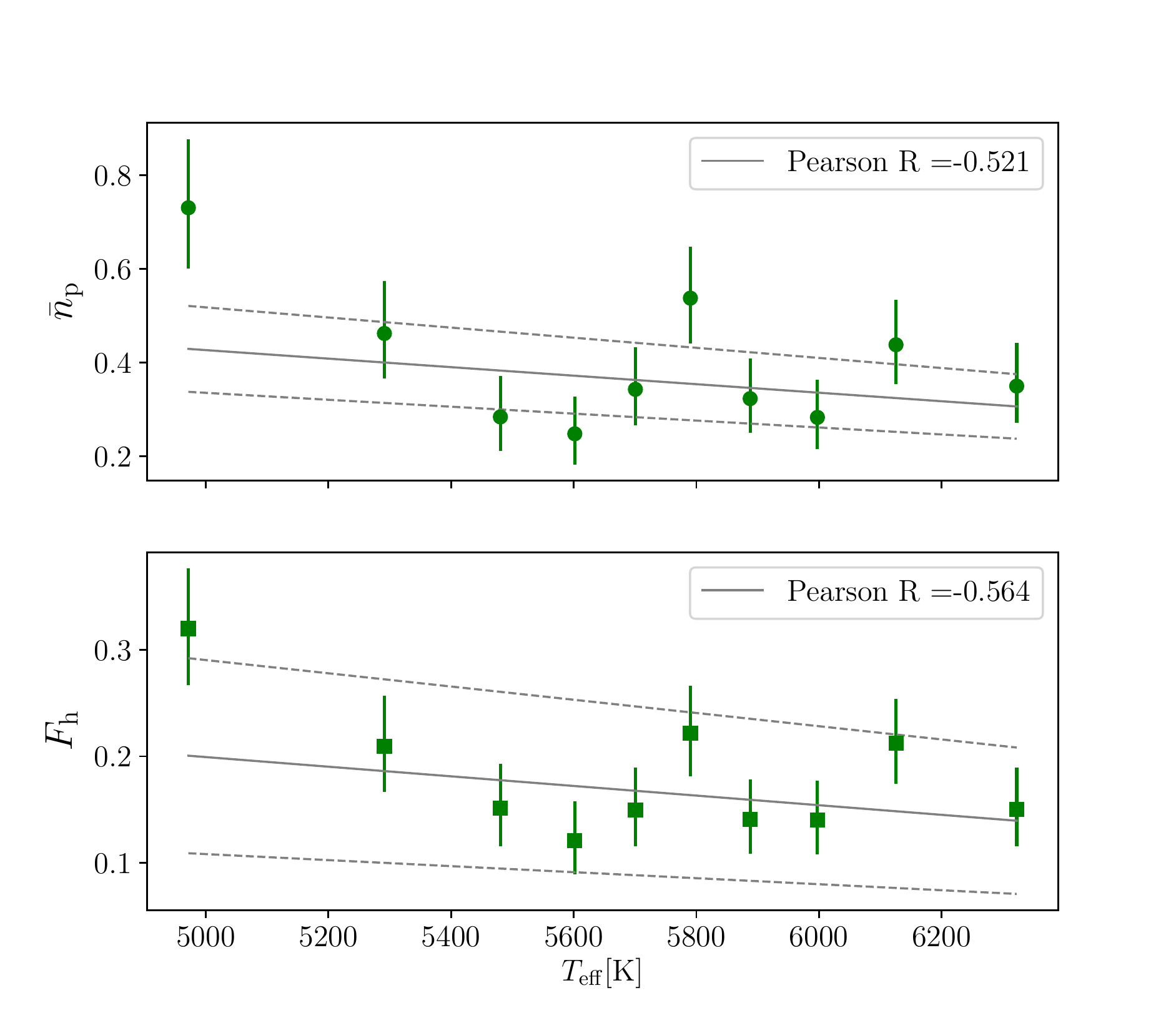}
\includegraphics[width=8.3cm]{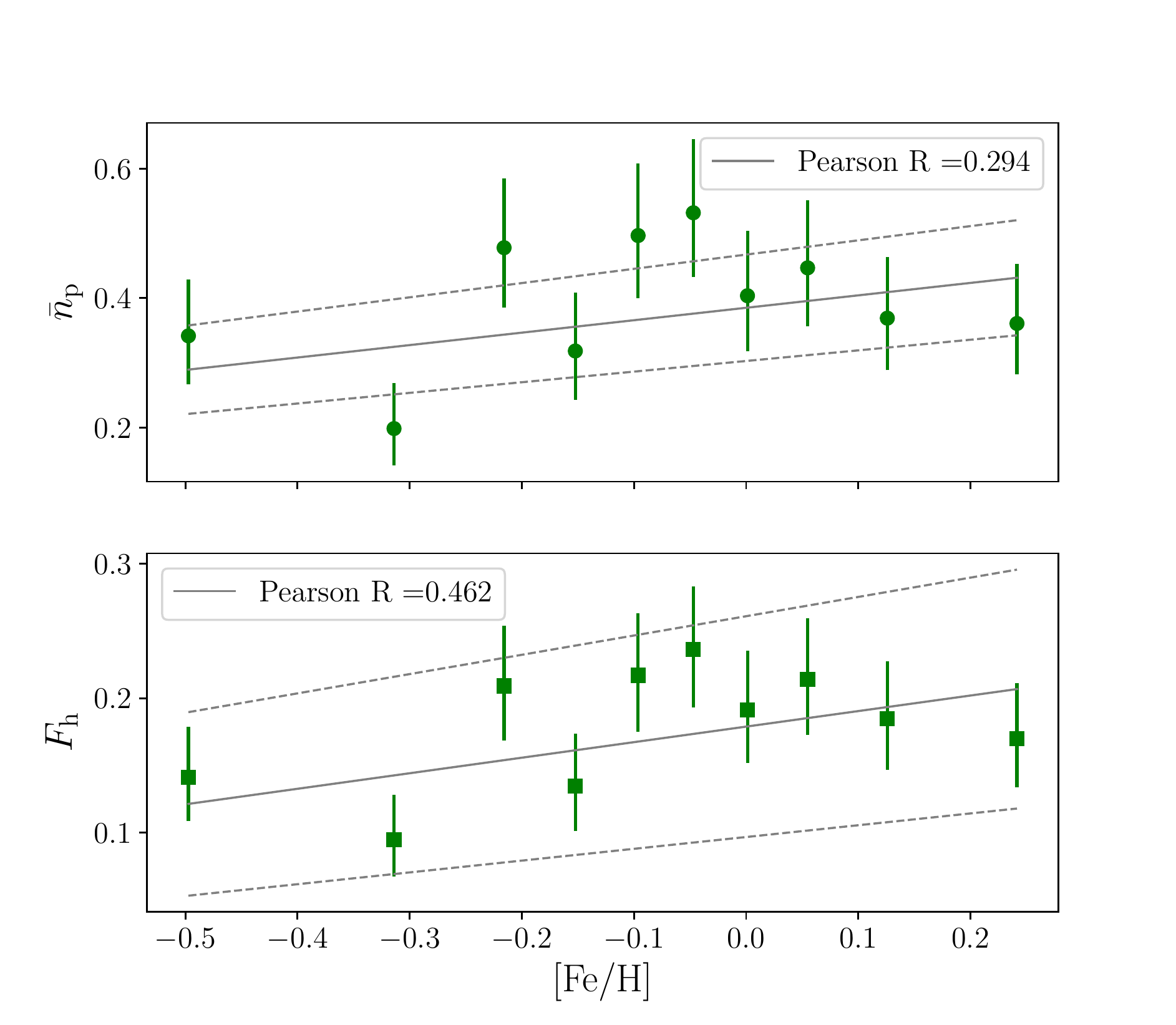}
\includegraphics[width=8.3cm]{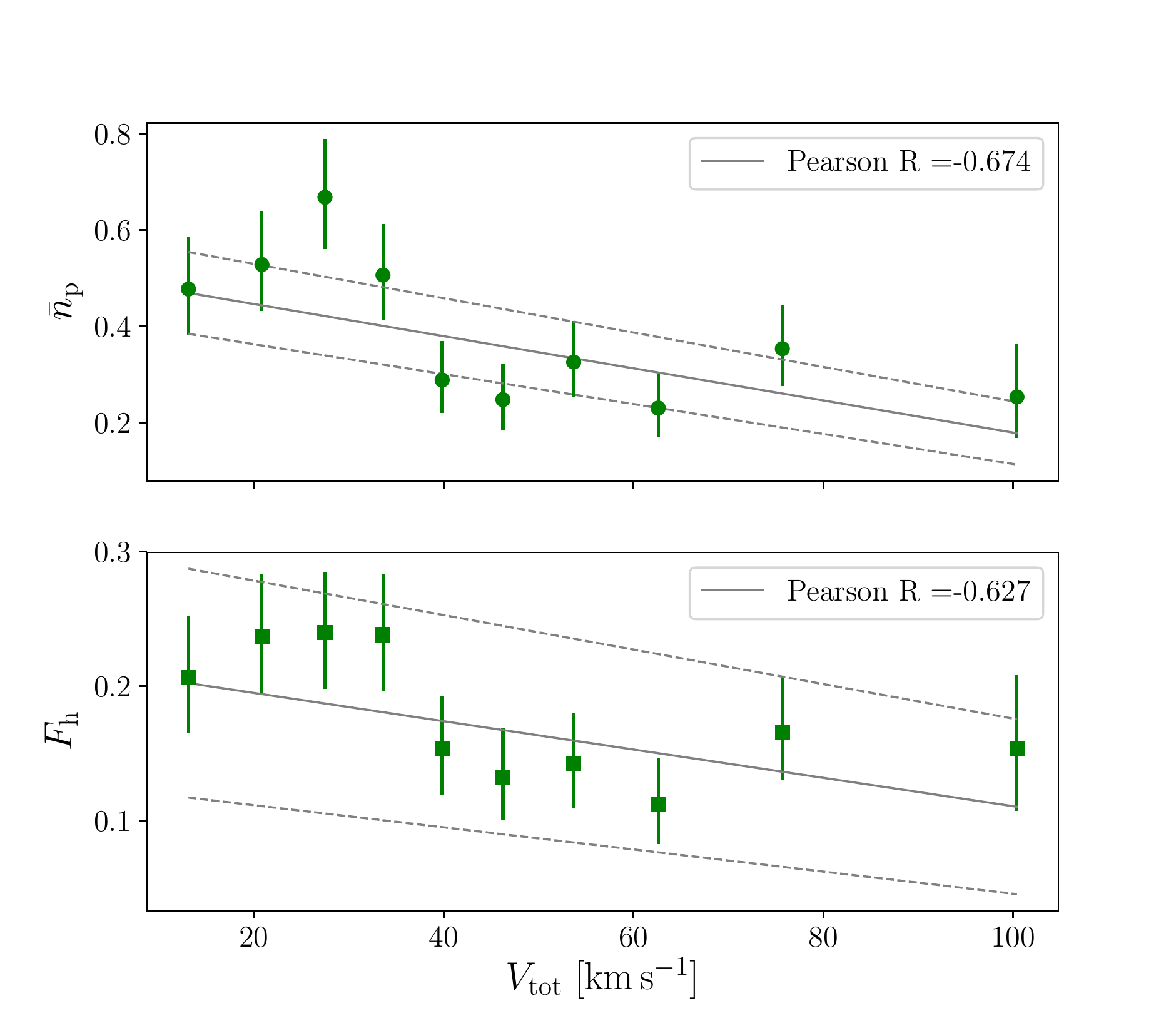}
\includegraphics[width=8.3cm]{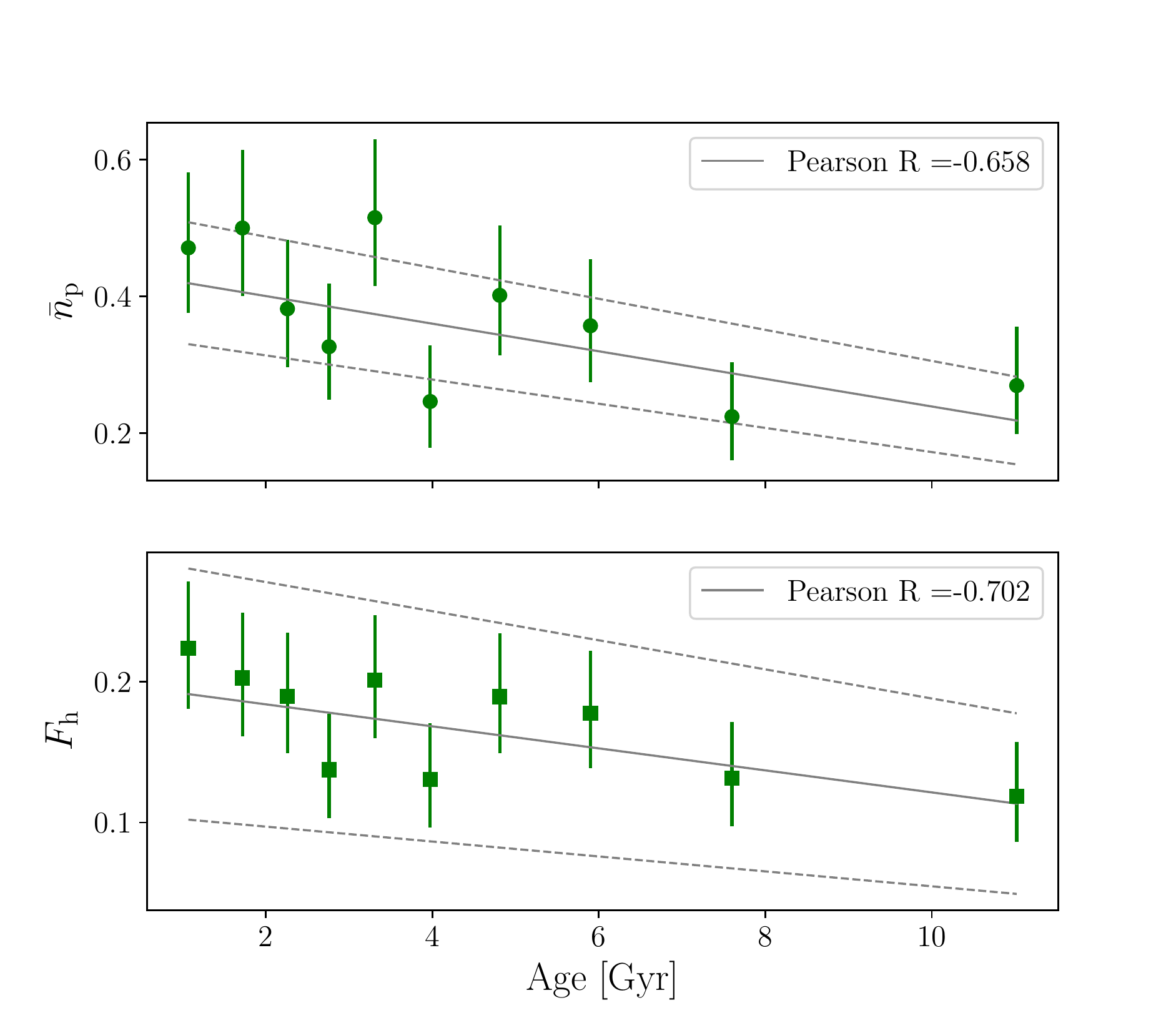}

\caption{
Dependence of close-in super-Earth occurrence rates ($\fpl$ and $\FplH$) on the stellar properties of effective temperature ($T_\mathrm{eff}$), metallicity ($[\mathrm{Fe}/\mathrm{H}]$), total velocity ($V_\mathrm{tot}$) and age.
The gray line represents best linear fit, with 1-$\sigma$ margins represented by the dotted lines. Also shown are the  Pearson correlation coefficients representing the significance of the linear relation.}
\label{figure:binsOcc}
\end{figure*}

For each parameter, we divided our sample to $10$ evenly sized bins. We focused mainly on the region of close-in super Earths, $R_{\mathrm{p}} = 1-2\,R_\oplus$, $P = 1-100$ days, as it exhibited the largest difference between the samples of the two Galactic populations. Figure~\ref{figure:binsOcc} shows the result, together with a simple linear fit (and lines corresponding to 1-$\sigma$ intervals of the fit parameters) as well the Pearson correlation coefficient, as a diagnostic for the direction and significance of the relationship. We tested these trends by dividing the sample into other set of bins and got similar results (not shown here). 

Our results regarding the correlations between occurrence rates of close-in super Earths and stellar spectral type and metallicity are consistent with those of previous works. We find an anti-correlation ($R_{\fpl} = -0.521$) with spectral type (i.e. small planets are more common around late-type stars) and a rather weak correlation with metallicity ($R_{\fpl} = 0.294$), which agrees with the results of \cite{Zhu19} and \cite{Bashietal20}. However, as most thin-disk stars in the solar neighbourhood are both more metal-rich and larger than thick-disk stars, these two effects should almost cancel each other when we estimate planet occurrence rate in the Galactic context. Nonetheless, when we examine the correlations with $V_{\mathrm{tot}}$ ($R_{\fpl} = -0.674$) and with age ($R_{\fpl} = -0.658$), an interesting trend emerges of increasing close-in super-Earth occurrence rates towards younger and slower stars (it is important to note here that stellar age and kinematics, in our cases approximated by $V_{\mathrm{tot}}$, are known to be correlated). These results suggest that stellar age and dynamical history might play an important role in shaping current small planet occurrence rates.

\section{Discussion}
\label{Discussion}

In this work we were able for the first time to estimate planet occurrence rates in different regimes of planetary properties using well defined samples of \textit{Kepler} stars in the Galactic context of the solar neighbourhood. Our results suggest that planets, and especially close-in super Earths, are more common around thin-disk stars compared to thick-disk stars. 

Stars in the different Galactic populations differ in terms of the stellar properties and their dynamical histories. It is difficult to study the details of the mutual relations between stellar properties and planet occurrence rates. Unfortunately, the current samples of the thick disk and the halo are too small, and we therefore could not study in detail planets that orbit fast stars, especially those affiliated with the stellar halo. Thus, future works should aim at extending this work, e.g. by using the \textit{K2} stellar sample, that spans a larger field of view and higher Galactic latitudes, and therefore contains more halo and thick-disk stars \citep[e.g.][]{Zinketal20}, and the TESS input catalogue \citep{Boleyetal21}.

Our results suggest to direct attention to stellar age and dynamical history as additional factors that may affect planet occurrence, and more specifically occurrence of small planets. It is plausible that as an older planetary system might have higher chances to have undergone some evolutionary processes that disrupted some of its planets, and especially the smaller ones that may be more easily affected by perturbations. This may help to explain the noticeable difference we found compared to close-in super Earths. Such processes may be the result of gravitational perturbations among the planets \citep{Mustilletal17} or the result of interaction with passing stars \citep{Kaibetal13, Winteretal20}. \cite{Adibekyanetal21}, after analysing the distribution of the orbital periods of planets detected by RV, found a suggestive difference between samples of stars from different Galactic populations and different age. They concluded that planetary architecture may be related to age, environment, or both, in agreement with our suggestion that those properties affect planet abundance across the Galaxy.

As opposed to iron content that is easier to deduce with good precision (as it has many absorption lines), it is harder to get accurate estimates of $[\alpha/\mathrm{Fe}]$. We therefore refrained from making any decisive claim regarding the details of the effect of stellar $\alpha$-enhancement on planets. However, we do suggest that $\alpha$-enhancement can serve more than just a proxy for age, but it might also be an important factor that could affect the types \citep{Santosetal17, Micheletal20} and occurrence rates of planets around thick-disk stars.

The study of exoplanets in the Galactic perspective is expanding in recent years. The exponentially increasing amount of available information both from \textit{Gaia}, ground based spectroscopic surveys and of course the constant flow of newly detected exoplanets, allow further exploration of the link between planets and stars in the Galaxy \citep[e.g.][]{BashiZucker19, Bashietal20, Chenetal21, Daietal21}. The growing sample of planetary systems around \textit{TESS} targets already takes shape \citep{Ganetal20, Weissetal21}, providing an opportunity to extend our analysis to quantify the difference between planetary system architectures \citep{BashiZucker21} in the Galactic context. It would be only natural to expect further discoveries in the next few years thanks to \textit{TESS} and the forthcoming \textit{PLATO} mission, aiming for $\sim10\%$ precision in stellar age determination for a large sample of stars \citep{Miglioetal17}. Furthermore, with the launch of The Nancy Grace Roman Space Telescope (\textit{Roman}), the future space mission of NASA that is expected to be launched in the next few years, new possibilities to explore exoplanets in the Galactic context will emerge. One of the primary goals of \textit{Roman} is to detect exoplanets using the microlensing method in the Galactic Bulge \citep{Spergeletal17}. It is estimated that \textit{Roman} will find about $54000$ microlensing events and will detect $1400$ planets \citep{Pennyetal19}, thus allowing an estimate of planet occurrence rates around the sub-population of bulge stars - a sub-population that is currently beyond the reach of exoplanet search programs.

\section*{Acknowledgements}

We thank the anonymous reviewer for the helpful comments and suggestions. This research was supported by the Israel Science Foundation (grant No. 848/16). We also acknowledge partial support by the Ministry of Science, Technology and Space, Israel. This paper includes data collected by the Kepler mission. Funding for the Kepler mission is provided by the NASA Science Mission directorate. This research has made use of the NASA Exoplanet Archive, which is operated by the California Institute of Technology, under contract with the National Aeronautics and Space Administration under the Exoplanet Exploration Program. This work has made use of data from the European Space Agency (ESA) mission Gaia (https://www.cosmos.esa.int/gaia), processed by the Gaia Data Processing and Analysis Consortium (DPAC, https://www.cosmos.esa.int/web/gaia/dpac/consortium). Funding for the DPAC has been provided by national institutions, in particular the institutions participating in the Gaia Multilateral Agreement. This paper also uses data from the LAMOST survey: Guoshoujing Telescope (the Large Sky Area Multi-Object Fiber Spectroscopic Telescope, LAMOST) is a National Major Scientific Project built by the Chinese Academy of Sciences. Funding for the project has been provided by the National Development and Reform Commission. LAMOST is operated and managed by the National Astronomical Observatories, Chinese Academy of Sciences.

\section*{Data Availability}

All data used in this work have been obtained from a cross-match using TOPCAT \citep{Taylor05} of publicly available databases. The data listed in Tables~\ref{table:KICsample} and \ref{table:KOIsample} is available online. All additional data used in this article will be shared on request by the corresponding author.


\bibliographystyle{mnras}



\appendix

\section{Effective number of stars}
\label{Appendix:EffectiveStars}


The probability to detect a transit is a product of three probabilities representing independent aspects of the transit detection process. The first one is the simple geometric transit probability $p_{\mathrm{tr}}$, i.e. the probability that the orbital inclination would allow transits. The second one encapsulates the detection efficiency $p_{\mathrm{det}}$, which strongly depends on the observational noise, including aspects of stellar variabiliy (red noise). The \textit{Kepler} detection statistic dubbed 'Multiple Event Statistic' \citep[MES;][]{Jenkins02, Christiansenetal12, Burkeetal15, Hsuetal19} is considered an adequate description the noise, and to first order, it can serve as a proxy to the transit signal-to-noise ratio (SNR), in order to estimate $p_{\mathrm{det}}$. The third one is related to the vetting process and the vetting probability $p_{\mathrm{vet}}$, i.e. the probability that a Threshold Crossing Event (TCE) is vetted to become a Kepler Object of Interest (KOI) and then assigned a planetary candidate disposition \citep{Thompsonetal18, Muldersetal18, Hsuetal19}.

We used the \texttt{KeplerPORTs} code \citep{BurkeCatanz17} to calculate, as a function of orbital period and planet radius, a per-target detection contour ($p_{\mathrm{tr}} \times p_{\mathrm{det}}$) for all $15881$ stars in our sample. 
We downloaded from the NASA Exoplanet Archive and \textit{Kepler} DR25 stellar data tables all the required input to \texttt{KeplerPORTs}, including the window function and one-sigma depth function data products \citep{BurkeCatanz17}, as well as the set of input values ($R_*$, \texttt{slpCDPPl}, \texttt{slpCDPPs}, duty cycle, data span, limb darkening coefficients). 
In our analysis we further used the recommendation of \cite{BurkeCatanz17} and ignored target stars reported to have deviant flux time series 
We used a grid based on $3500$ evenly spaced trial orbital periods in the range $P=1-400$ days and $1000$ evenly spaced (logarithmically) trial planet radii in the range $R_{\mathrm{p}} = 0.5-15\,R_\oplus$. 

As for the vetting process, we followed \cite{Yangetal20} and used the KOI vetting efficiency fit of \cite{Muldersetal18} for disposition scores larger than $0.9$:

\begin{equation}
    p_{\mathrm{vet}} = 0.63 R_{\mathrm{p}}^{0.19} \begin{cases}
    (P/P_{\mathrm{break}})^{-0.07}, & \text{if $P < P_{\mathrm{break}}$}\\
    (P/P_{\mathrm{break}})^{-0.39}, & \text{otherwise} \, ,
  \end{cases}
\end{equation}
with $P_{\mathrm{break}} = 53$ days. 

Thus, we approximated the probability $p_{\mathrm{i,j}}$ to detected planet $j$ around star $i$ by:
\begin{equation}
\label{eq: detLim}
    p_{\mathrm{\mathrm{i,j}}}=p_{\mathrm{tr}}\times p_{\mathrm{det}} \times p_{\mathrm{\mathrm{vet}}}\, ,
\end{equation}


In order to estimate the effective number of stars searched, we used the following prescription: 

(i) For $\Neffn$, the effective number of stars searched, which we later used to estimate the average number of planets per star, we selected for each star the pre-calculated detection probability matrix. We then simulated $N = 1000$ random planets with relevant properties drawn from the empirical detected planet distributions of the planet parameters $P$ and $R_{\mathrm{p}}$ (and their uncertainties) as discussed in subsection \ref{Planetsample}. In the process of drawing the $N$ simulated planets, the probability to draw a planet was inversely proportional to the probability to detect it according to the averaged completeness maps of all KIC stars of our sample. Thus, planets with lower detection probability will have a higher probability to be produced in the simulation. 
Consequently, the probability to detect these simulated planets is simply the average value of:
\begin{equation}
    p_\mathrm{i} = \frac{1}{N}  \sum_{\mathrm{j=1}}^{N}  p_{\mathrm{i,j}} \, ,
\end{equation}
The expected value for the effective number of all stars searched, $\Neffn$, is simply the sum of these probabilities over all target stars $N_{\mathrm{sample}}$:
\begin{equation}
    \Neffn = \sum_{\mathrm{i=1}}^{N_{\mathrm{sample}}} p_{\mathrm{i}} \, ,
\end{equation}


(ii) As for $\NeffF$, the effective number of stars searched, that we used in order to estimate the fraction of stars with planets, we used a similar procedure with, some small changes. Instead of simulating $N = 1000$ random planets, we simulated $N = 1000$ planet-host stars, based on the empirical sample of planet host stars. In the process of drawing the $N$ simulated planet-host stars, the probability to draw a planetary system was inversely proportional to the planetary multiplicity distribution times the sum of probabilities to detect its planets according to the averaged completeness maps of all KIC stars of our sample. Thus, we preserved in our calculations the overall multiplicity distribution of the planet-host sample. To first order, the probability to detect exactly $N_\mathrm{p}$ planets is the sum of the probabilities to detect each individual planet separately (indeed, in principle one needs to subtract from this number the conditional probabilities to detect each planet given the others, yet these probabilities are negligible). Thus, the probability to detect these planetary systems is simply the average value of: 
\begin{equation}
    \tilde{p_\mathrm{i}} = \frac{1}{N}  \sum_{\mathrm{k=1}}^{N} \sum_{\mathrm{j=1}}^{(N_\mathrm{p})_\mathrm{k}}p_{\mathrm{i,j}} \, ,
\end{equation}
The expected value for the effective number of all stars searched, $\NeffF$, is simply the sum of these probabilities over all $N_{\mathrm{sample}}$ target stars :
\begin{equation}
    \NeffF = \sum_{\mathrm{i=1}}^{N_{\mathrm{sample}}} \tilde{p_\mathrm{i}} \, ,
\end{equation}


\bsp	
\label{lastpage}
\end{document}